\documentclass[preprint,tightenlines,nofootinbib,superscriptaddress,floatfix]{revtex4}

\usepackage[bookmarks=false,hyperfootnotes=false]{hyperref}
\usepackage{graphicx}
\usepackage{amsmath}
\usepackage{amssymb}
\usepackage{xspace}

\usepackage{calc}

\newcommand{\eq}{=}

\newcommand{\ub}{\text{ub}}

\newcommand{\nul}{\nu_l}

\newcommand{\bea}{\begin{align}}
\newcommand{\eea}{\end{align}}
\newcommand{\bsea}{\begin{subequations}}
\newcommand{\esea}{\end{subequations}}

\newcommand{\babar}{\mbox{\ensuremath{{\displaystyle B}\!{\scriptstyle A}{\displaystyle B}\!{\scriptstyle AR}}}\xspace}
\newcommand{\belle}{Belle\xspace}

\allowdisplaybreaks[2]

\newcommand{\mytable}[3]{
  \begin{table}[#1]
    \begin{center}
      #2\\\mytabcaption{\textwidth}{#3}
    \end{center}
  \end{table}
}

\newcommand{\myfigure}[3]{
  \begin{figure}[#1]
    \begin{center}
      #2\\\myfigcaption{\textwidth}{#3}
    \end{center}
  \end{figure}
}

\begin{document}

%%%%%%%%%%%%%%%%%%%%%%%%%%%%%%%%%%%%%%%%%%%%%%%%%%%%%%%%%%%%%%%%%%%%%%%%%%%%%%%%
% Title page
%%%%%%%%%%%%%%%%%%%%%%%%%%%%%%%%%%%%%%%%%%%%%%%%%%%%%%%%%%%%%%%%%%%%%%%%%%%%%%%%

\preprint{\vbox{\hbox{HU-EP-11/30}\hbox{\today}}}

\title{Semileptonic decays and the determination of $\left| V_{\text{ub}} \right|$}

\author{Florian U. Bernlochner}
\affiliation{Humboldt University of Berlin, 12489 Berlin, Germany \\ \& \\ University of Victoria,  V8Y1L1 British Columbia, Canada  }

\collaboration{On behalf of the \babar Collaboration}

\vspace{3ex}

%%%%%%%%%%%%%%%%%%%%%%%%%%%%%%%%%%%%%%%%%%%%%%%%%%%%%%%%%%%%%%%%%%%%%%%%%%%%%%%%
\begin{abstract}

I present an overview of the experimental and theoretical situation for the determination of  $\left| V_{\text{ub}} \right|$ from semileptonic $B$-meson decays. 

\vfill
\noindent\emph{Proceedings of the conference of Flavor Physics and CP Violation 2011\\
Kibbutz Maale Hachamisha, Israel, May 23-27 2011}

\end{abstract}
%%%%%%%%%%%%%%%%%%%%%%%%%%%%%%%%%%%%%%%%%%%%%%%%%%%%%%%%%%%%%%%%%%%%%%%%%%%%%%%%

\maketitle

\section{\boldmath Introduction}
\label{sec:introduction}
%%%%%%%%%%%%%%%%%%%%%%%%%%%%%%%%%%%%%%%%%%%%%%%%%%%%%%%%%%%%%%%%%%%%%%%%%%%%%%%%

The precision determination of the Cabibbo-Kobayashi-Maskawa (CKM) matrix element $\left| V_{\text{ub}} \right|$ is of particular importance to test the Yukawa sector of the Standard Model (SM). More precisely, it's needed to test the mechanism which explains the occurrence of \emph{Charge} and \emph{Parity} (CP) violating effects in weak decays. Such effects are caused by the presence of an irreducible complex phase in the unitary $3 \times 3$ CKM matrix which can be illustrated by using one of the triangle equation derived from the unitarity constraint of the CKM matrix, i.e. $V_{\text{ud}} \, V_{\text{ub}}^* + V_{\text{cd}} \, V_{\text{cb}}^* + V_{\text{td}} \, V_{\text{tb}}^* = 0$. Given the proper normalization of the sides of the triangle the complex phase corresponds to the apex of the so-called \emph{unitary} triangle in the complex plane, what is illustrated in Fig. \ref{fig:unitary_triangle}. The left side of the unitary triangle is proportional to the absolute value of $V_{\text{ub}}$, which can be combined with the direct measurements of the angles of the triangle and the other CKM matrix elements to test the predicted unitarity of the SM CKM matrix. Such tests were performed e.g. by the authors of Refs.~\cite{Hocker:2001xe} and \cite{Ciuchini:2003rk} and found an excellent agreement with unitarity within todays experimental precision. 

The CKM matrix element $\left| V_{\text{ub}} \right|$ can be determined from a multitude of weak decays which involve either inclusive or exclusive final states and exhibit different experimental or theoretical challenges. Fig.~\ref{fig:had_semlept_lep_dec} illustrates the leptonic, hadronic, and semileptonic transitions whose partial or total decay rates are proportional to $\backsim \left| V_{\text{ub}} \right|^2$. In this presentation I will focus on the determination of $\left| V_{\text{ub}} \right|$ from semileptonic decays, which have some definite advantages over the determination of $\left| V_{\text{ub}} \right|$ from leptonic, e.g. through $B \to \tau \bar \nu_\tau$, and hadronic decays: The leptonic determination is experimentally challenging, and the prediction of the hadronic decay rate involves complicated non-perturbative and perturbative corrections due to the strong interaction processes between the two hadronic final states. The study of semileptonic decays offers some middle ground between experimental and theoretical challenges: the presence of a high-energetic lepton offers a good discriminator from other weak decays, and the determination of the decay rate is simplified due to the factorization of the matrix element into a hadronic and leptonic current. In practice, however, the presence of the much more abundant semileptonic $b \to c$ transition complicates things and one neither has to identify one particular exclusive final state (e.g. one pion), restrict the measurement to regions of phase-space where charmed semileptonic decays are kinematically suppressed or forbidden, or use data mining algorithms building on multivariate methods which can separate charmed and charmless semileptonic transitions. The presented results in this talk determine $\left| V_{\text{ub}} \right|$ either via the reconstruction of an exclusive final state, or measure the fully inclusive (partial) decay rate.

\myfigure{tbp!}{
\vspace{4ex}
 \includegraphics[width=0.42\textwidth]{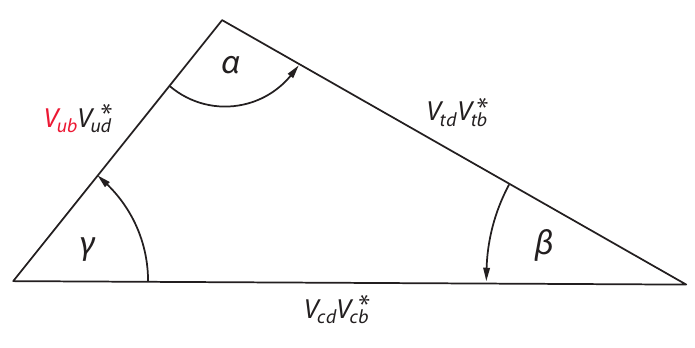}
 }{ The unitary triangle for  $V_{\text{ud}} \, V_{\text{ub}}^* + V_{\text{cd}} \, V_{\text{cb}}^* + V_{\text{td}} \, V_{\text{tb}}^* = 0$ is shown. The illustration was taken from Ref.~\cite{doi:10.1146/annurev.nucl.012809.104421}. \label{fig:unitary_triangle} }

 \myfigure{tbp!}{
 \includegraphics[width=0.82\textwidth]{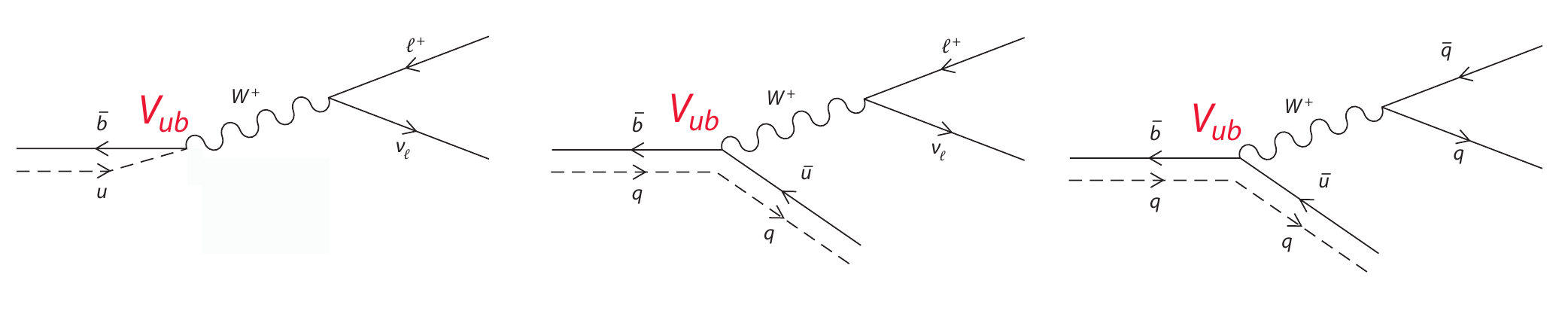}
 }{ The Feynman graphs for the leptonic, semileptonic, and hadronic decays of a $B$ meson are shown. The illustrations were taken from Ref.~\cite{doi:10.1146/annurev.nucl.012809.104421}. \label{fig:had_semlept_lep_dec} }

%%%%%%%%%%%%%%%%%%%%%%%%%%%%%%%%%%%%%%%%%%%%%%%%%%%%%%%%%%%%%%%%%%%%%%%%%%%%%%%%
\section{\boldmath Inclusive determination of $\left| V_{ub} \right|$}
%%%%%%%%%%%%%%%%%%%%%%%%%%%%%%%%%%%%%%%%%%%%%%%%%%%%%%%%%%%%%%%%%%%%%%%%%%%%%%%%

For inclusive $B \to X_u \, l \, \bar \nu_l$ decays, where $X_u$ denotes a hadronic system consisting of one or more mesons with at least one $u$ quark produced in the weak $b \to u$ quark transition, a prediction for the total decay rate can be readily obtained in the heavy quark expansion (HQE) of the Standard Model Lagrangian. The semileptonic transition can be expressed as a sum of local operators using an operator product expansion (OPE), cf. e.g. Ref.~\cite{Manohar:2000dt} for a review. Unfortunately the dominant $B \to X_c \, l \, \bar \nu_l$ decays do not allow the direct measurement of the total decay rate. Many analyses are thus forced thus to measure the partial branching fraction in a region of phase-space, where the background from $b \to c$ transitions is considerably suppressed or kinematically forbidden. Although in these regions a clear separation from charmed semileptonic decays emerge, the  OPE breaks down and non-perturbative and perturbative correction to the partial decay rate become very important. In particular, the Fermi motion of the $b$ quark inside the $B$ meson becomes a relevant factor which e.g. determines the shape of the lepton momentum spectrum near the endpoint or the hadronic invariant mass distribution, $m_X$, for small values of $m_X$. This behaviour of the OPE can be remedied by summing the most divergent contributions of the expansion into a single function, the so-called \emph{shape function}. In practice, the functional form of the shape function is unknown. Its moments, however, can be related to the moments of the measured mass and lepton spectra of $b \to c$ decays and observables from $b \to s \gamma$ decays, cf. e.g. Refs.~\cite{Neubert:1993um}. Thus the predictions of the partial decay rates of Refs.~\cite{Lange:2005yw,Gardi:2008bb,Gambino:2007rp,Aglietti:2007ik} employ a variety of model functions to describe the shape function. The first few moments of these model function are matched to the experimental available input, and are used under the premise that the precise functional form of the shape function only has a negligible impact on the resulting partial decay rate. Experimentally, the use of multivariate methods allowed reducing the dependence of the shape function by measuring a large fraction of the total phase space. In this talk the recent measurements of Refs.~\cite{PhysRevLett.104.021801}  is reviewed whose authors successfully employ such a multivariate technique to measure about $\backsim 90\%$ of the $B \to X_u \, l \, \bar \nu_l$ phase-spac. Furthermore, I review the recent analysis of Ref.~\cite{babarincl}, whose authors use a veto based approach and employ phase-space cuts to reject charmed background. 

%%%%%%%%%%%%%%%%%%%%%%%%%%%%%%%%%%%%%%%%%%%%%%%%%%%%%%%%%%%%%%%%%%%%%%%%%%%%%%%%
\subsection{\boldmath Study of $B \to X_u \, l \, \bar \nu_l$ from \belle: Ref.~\cite{PhysRevLett.104.021801} }
\label{sec:belleincl}
%%%%%%%%%%%%%%%%%%%%%%%%%%%%%%%%%%%%%%%%%%%%%%%%%%%%%%%%%%%%%%%%%%%%%%%%%%%%%%%%

The authors of Ref.~\cite{PhysRevLett.104.021801} measure the partial $B \to X_u \, l \, \bar \nu_l$ branching fraction with a non-precedented coverage of the allowed $B \to X_u \, l \, \bar \nu_l$ phase-space. In total $657 \times 10^6$ $B \bar B$ pairs, measured at the KEK-II $B$ factory with the \belle detector, were analyzed in a tagged approach, i.e. one of the decaying $B$ mesons is completely reconstructed to infer the kinematics and flavor for the recoiling $B$ meson. This so-called tag or reco $B$-meson is reconstructed using a multitude of exclusive hadronic decay modes. For each such tag $B$ meson candidate the beam-energy substituted mass
\begin{align}\label{eq:mes}
 m_{bc/ES} & \eq \sqrt{ s/4 - \left| \vec p_B^{\,\,*} \right|^2 } \, , 
\end{align}
and the energy difference 
\begin{align}\label{eq:deltaE}
 \Delta E = E_B^* - \sqrt{s}/2 
\end{align}
is reconstructed, where $\left( E_B^*, \vec p_B^{\,\,*}\right)$ is the four-momentum of the tagged $B$-meson candidate in the $\Upsilon(4S)$ rest frame, and $\sqrt{s}$ denotes the beam energy. For correctly reconstructed reco $B$-meson candidates one expects $m_{bc/ES}$ to peak at the $B$-meson mass, and $\Delta E$ at zero. The primary vertex of the $B$-meson reco candidates is determined using a vertex fit and in if more than one candidate is present in an event, only the one with the highest significance with respect to the quality of the fit is retained. Further, a quality cut on the $\chi^2$ value of the vertex fit is implemented. The authors require a reco candidate $B$-meson to lie within $ 5.27\, \text{GeV} < m_{bc/ES} < 5.29\, \text{GeV} $ and $ - 0.05\, \text{GeV} < \Delta E < 0.05\, \text{GeV}$. The shape of combinatorial background in the given $ m_{bc/ES}$ and $\Delta E$ window is estimated from Monte Carlo (MC) simulations and subtracted from the candidates. Contributions from non-$B\bar B$ continuum background is subtracted by a off-resonance data sample which was taken below the $\Upsilon(4S)$ resonance scaled by the integrated on- and off-resonance luminosity ratio. Electron and muon candidates from the recoiling other $B$ meson are required to pass angular acceptance cuts and to be identified by a particle identification algorithm. Due to the reconstructed $B$-meson four-momentum from the tag side, the lepton three momentum can be boosted into the rest frame of the recoiling $B$-meson. The hadronic $X_u$ system associated with the $B \to X_u \, l \, \bar \nu_l$ decay is reconstructed from charged tracks and energy depositions in the calorimeter that are not associated with the tagged side of the event. The further $B \to X_u \, l \, \bar \nu_l$ signal selection is based on a non-linear multivariate boosted decision tree (BDT), cf. Ref.~\cite{bdt}. The BDT incorporates a total of 17 discriminating variables to create a single classifier, which separate the $B \to X_u \, l \, \bar \nu_l$ semileptonic decays from other background. These discriminative values include non-exhaustively kinematic quantities, the number of identified kaons in an event, $m_{bc/ES}$, the absolute value of the net charge of the event, and the track multiplicity, which tends to be higher for recoil candidates originating from $b \to c$ transitions. Further, if a low-momentum pion is present in a given event it is associated with a potential strong $D^* \to D \pi$ decay. Due to the low momentum transfer of the $D^*$ to the $D$ system, the pion three-momentum carries almost all of the three-momentum of the $D^*$ and can be used to infer the $D^*$ four-momentum. This can be used to calculate the missing mass squared associated with a $B \to D^* \, l \, \bar \nu_l$ decay, i.e. 
\begin{align}\label{eq:mmsqDstar}
m_{\text{miss} \, D^*}^2 & \eq \left( p_{B'} - p_{D^*} - p_l  \right)^2 \, ,
\end{align}
where $p_{B'}$ is the four-momentum of the recoiling signal $B$-meson inferred from the tagged side, $p_{D^*}$ denotes the four-momentum inferred from the slow pion, and $p_l$ is the four-momentum of the lepton candidate. In case of a true $B \to D^* \, l \, \bar \nu_l$ decay, Eq.~ \ref{eq:mmsqDstar} is expected to peak at zero. Using these discriminators as input for the BDT a good signal to background separation can be obtained down to a lower lepton momentum cut in the $B$-meson rest frame, $\left| \vec p_l^{\,\,*B} \right|$, of  $1$ GeV/$c^2$. This allows to measure the partial branching fraction of of about $90\%$ of the allowed phase-space of the $B \to X_u \, l \, \bar \nu_l$ transition. 

The candidates passing the preselection and the selection of the BDT classifier are analyzed in a two-dimensional fit in the reconstructed hadronic invariant mass of the $X_u$ system, $m_X$, and the four-momentum transfer squared of the $B$-meson to the $X_u$ system, i.e. in the $\Upsilon(4S)$ rest frame
\begin{align}\label{eq:q2_tagged}
 q^2 & = \left( (\sqrt{s}, \vec 0\,) - p_B^* - p_X^* \right)^2,
\end{align}
 where $p_B^*$ denotes the four-momentum of the tagged $B$ meson, and $p_X^*$ is the four-momentum of the reconstructed hadronic $X_u$ system. The free parameters in the two-dimensional fit are the yields of $B \to X_u \, l \, \bar \nu_l$, $B \to X_c \, l \, \bar \nu_l$, and other backgrounds caused mainly by secondary and misidentified leptons.

\myfigure{tbpH!}{
\vspace{-3.5ex} \hspace{-11ex}
\parbox{0.40\textwidth} {   \includegraphics[width=0.70\textwidth]{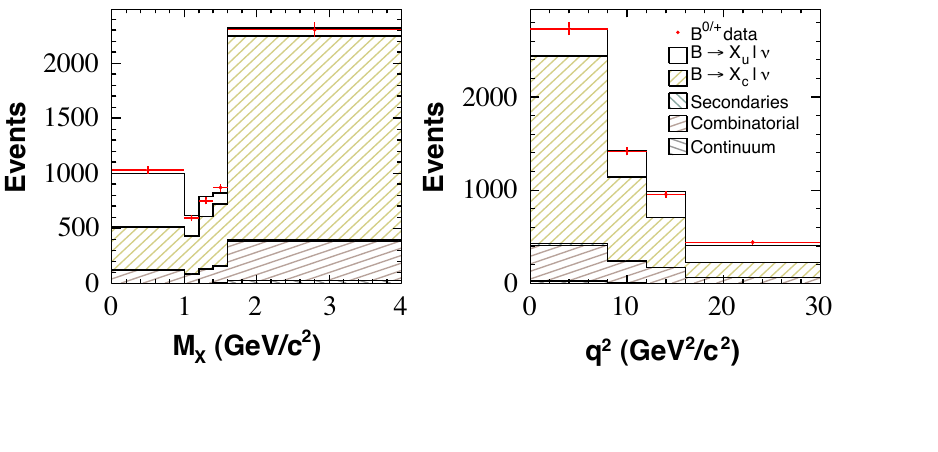}    } 
\vspace{-3ex}
}{$\qquad$ The one-dimensional projections of the fitted $m_X$ and $q^2$ distributions of Ref.~\cite{PhysRevLett.104.021801} are shown. \label{fig_mxq2_belle}    }

The $B \to X_u \, l \, \bar \nu_l$ contributions are modelled using a hybrid mix of inclusive and exclusive contributions. Resonant contributions from $B \to \pi \, l \, \bar \nu_l$, $B \to \rho \, l \, \bar \nu_l$, $B \to \omega \, l \, \bar \nu_l$ are modelled using the form factor predictions of Ref.~\cite{Ball:hep-ph0406232,Ball:2004rg}. The resonant decays into $B \to \eta \, l \, \bar \nu_l$ and  $B \to \eta' \, l \, \bar \nu_l$ are modelled using the quark-potential model of Ref.~\cite{Scora:1995ty}. The corresponding exclusive branching fractions are fixed at the measured world-averages calculated by Ref.~\cite{Barberio:2008fa}. The non-resonant contributions are modelled using the shape function parametrization of Ref.~\cite{1126-6708-1999-06-017}. The hybrid MC is required to match the moments of the $q^2$ and $m_X$ distributions predicted by Ref.~\cite{Gambino:2007rp}.

The $B \to X_c \, l \, \bar \nu_l$ contributions are modelled as the sum of the exclusive $B \to D \, l \, \bar \nu_l$, $B \to D^* \, l \, \bar \nu_l$, $B \to D^{**} \, l \, \bar \nu_l$, and non-resonant $B \to D^{(*)} \, \pi \, l \, \bar \nu_l$ decays modelled after Refs.~\cite{Caprini:1997mu,Leibovich:1997em,Goity:1994xn}. The non-perturbative shape parameters of the $B \to D^{(*)} \, l \, \bar \nu_l$ decays are fixed at the measured world average of Ref.~\cite{Barberio:2008fa}, and the branching fractions are fixed at the measured values of Ref.~\cite{Amsler:2008zz}. The narrow  $B \to D^{**} \, l \, \bar \nu_l$ and non-resonant  $B \to D^{(*)} \, \pi \, l \, \bar \nu_l$ branching fractions are fixed at the averaged values of Ref.~\cite{Barberio:2008fa}. The broad $B \to D^{**} \, l \, \bar \nu_l$ branching fractions are fixed to fill the gap between the inclusive and summed exclusive branching fractions. 

For the other background components the authors rely on the MC prediction of the differential shapes in $q^2$ and $m_X$. 

The one-dimensional projections of the fitted and measured $m_X$-$q^2$ distribution are depicted in Fig.~\ref{fig_mxq2_belle} and the fit has a good statistical significance of $12\%$. The binning was chosen in order to minimize the dependence on the $B \to X_u \, l \, \bar \nu_l$ signal model. The partial branching fraction is calculated as
\begin{align}\label{eq:pbf_belle_incl}
  \Delta \mathcal{B}\left(\left| \vec p_l^{\,\,*B} \right|> 1.0 \, \text{GeV}\right) & = \tfrac{ N_{b\to u}^\Delta }{2 \epsilon_{b \to u}^\Delta \, N_{\text{tag}} } \times \left( 1 - \delta_{\text{QED}} \right)   \, ,
\end{align}
where $ N_{b\to u}^\Delta $ and $ \epsilon_{b \to u}^\Delta$ correspond to the fitted $B \to X_u \, l \, \bar \nu_l$ signal yield and efficiency, and $ N_{\text{tag}}$ corresponds to the number of tagged $B$ decays. The factor $\delta_{\text{QED}}$ corresponds the arising QED correction which is predicted using the algorithm of Ref.~\cite{Barberio:1993qi}. The systematic uncertainties to the partial decay rate were studied by varying the constitution of the $B \to X_u \, l \, \bar \nu_l$ hybrid mix, the shape function moments, and the $B \to X_c \, l \, \bar \nu_l$ contributions. The value of $\left| V_{\text{ub}} \right|$ can be readily obtained from Eq.~(\ref{eq:pbf_belle_incl}) using the predictions for the partial differential decay rates of Refs.~\cite{Lange:2005yw,Gardi:2008bb,Gambino:2007rp} via 
\begin{align}\label{eq:vub}
 \left| V_{\text{ub}} \right| & \eq \sqrt{ \tfrac{ \Delta \mathcal{B}(p_l > 1.0) } { \tau_B \, \Delta \zeta(p_l^* > 1.0) }  } \, 
\end{align}
where $\tau_B$ denotes the averaged $B^+$ and $B^0$-meson life time, and $\Delta \zeta$ the prediction for the partial decay rate stripped of the squared CKM matrix element. The determined values for $ \left| V_{\text{ub}} \right| $ are given in Table \ref{vub:belle_incl}.

 \mytable{tbH!}{
{\footnotesize 
  \begin{tabular}{l | c c c} 
   &  BLNP \cite{Lange:2005yw} & GGOU \cite{Gambino:2007rp} & DGE  \cite{Gardi:2008bb} \\  \hline
    $ \left| V_{\text{ub}} \right|  \times 10^{3}$ &   ${\bf 4.37 \pm 0.26 {}^{+0.23}_{-0.21} }$  &  ${\bf 4.41 \pm 0.26 {}^{+0.12}_{-0.22} }$ &  ${\bf 4.46 \pm 0.26  {}^{+0.15}_{-0.16}  }$
  \end{tabular}\hspace{5mm}\vspace{2ex}
 }
} { The determined values of $\left| V_{\text{ub}} \right|$ of Ref.~\cite{PhysRevLett.104.021801} are listed using the prediction of the partial decay rates of Refs.~\cite{Lange:2005yw,Gardi:2008bb,Gambino:2007rp}. The uncertainties are experimental and due to theory. \label{vub:belle_incl}  }  

%%%%%%%%%%%%%%%%%%%%%%%%%%%%%%%%%%%%%%%%%%%%%%%%%%%%%%%%%%%%%%%%%%%%%%%%%%%%%%%%
\subsection{\boldmath Study of $B \to X_u \, l \, \bar \nu_l$ from \babar:  Ref.~\cite{babarincl} }
\label{sec:belleincl}
%%%%%%%%%%%%%%%%%%%%%%%%%%%%%%%%%%%%%%%%%%%%%%%%%%%%%%%%%%%%%%%%%%%%%%%%%%%%%%%%

The authors of Ref.~\cite{babarincl} measure the partial branching fraction of $B \to X_u \, l \, \bar \nu_l$ using various kinematic cuts to suppress the predominant $B \to X_c \, l \, \bar \nu_l$ background. In particular, cuts on the following kinematic variables are studied to separate signal from background contributions: the four-momentum transfer of the $B$ meson to the $X_u$ system squared, $q^2$, the lepton three-momentum in the $B$-meson rest frame, $p_l$, the light-cone momentum of the hadronic system obtained the $B$-meson rest frame, $p^+ = E_X - \left|\vec p_X \right|$, and the hadronic invariant mass, $m_X$. In total $467 \times 10^6$ $B \bar B$ decays measured at the PEP-II $B$ factory with the \babar detector were analyzed using a tagged approach. The recoiling $B$ meson is reconstructed using a semi-exclusive algorithm based on hadronic $B \to D^{(*)} \, Y$ decays, where $Y$ is a charged system composed of pions and kaons. The kinematic consistency of the $B$ meson candidate is checked using the beam-energy substituted mass Eq.~(\ref{eq:mes}) and the energy difference Eq.~(\ref{eq:deltaE}). The selection requires a value of $\Delta E$ compatible with zero within about three standard deviations of the resolution uncertainty. If more than one $B$ meson candidate is present in an event, the candidate with the lowest $\chi^2$ from the summed $\chi^2$ terms of a kinematic vertex fit, the central value of the reconstructed $D$ mass, and the compatibility of $\Delta E$ with zero, is chosen. This selection results in at least one $B$ meson candidate in $0.3\%$ of all $B^0 \bar B^0$ and in $0.5\%$ of all $B^+ B^-$ events. 

The $B \to X_u \, l \, \bar \nu_l$ signal selection requires at least one electron or muon candidate, which is identified using a particle identification algorithm, and needs to pass an angular acceptance cut. A lower cut on the three-momentum in the $B$-meson rest frame, $\left| \vec p_l^{\,\,*B} \right|$, of $1$ GeV/$c^2$ is imposed in order to suppress background from cascade and $\tau$ decays. Muon candidates, which are consistent with originating from a $J/\psi$ when paired with another charged track of opposite charge in the event, are rejected. Further, electrons which are consistent with originating from a $\gamma \to e^+ e^-$ conversion when paired with another charged track of opposite charge in the event, are rejected. The hadronic system $X$ is reconstructed from charged tracks and energy depositions in the calorimeter that are not associated with the tagged $B$-meson candidate. Requiring exactly one electron or muon candidate suppresses $B \to X_c \, l \, \bar \nu_l$ background which frequently produce a second lepton in cascade decays. In addition, a charge correlation cut with the determined charge of the $B$ meson from the tag side, $Q_b$, and the charge of the reconstructed lepton, $Q_l$ is imposed, such that, $Q_b \, Q_l < 0$. In addition, the total charge of the event is required to be zero, i.e. $Q_{\text{tot}} = Q_b+ Q_l + Q_X = 0$, where $Q_X$ is the charge of the hadronic $X$ system. After this selection three main sources of background remain: combinatorial background from the tagged side; background from $B \to X_c \, l \, \bar \nu_l$ and cascades; background from $B \to X_u \, \tau \, \bar \nu_l$ decays with $\tau \to e$ or $\mu$. The combinatorial background is subtracted on the tag side by a maximum likelihood fit of the $m_{ES}$ distribution. The other two sources can be reduced by reconstructing the missing mass from the missing four-momentum of the event, i.e. in the $\Upsilon(4S)$ rest frame
\begin{align}
 p_{\text{miss}}^* & \eq \left( \sqrt{s}, \vec 0 \right) - p_B^* - p_X^* - p_l^* \, , 
\end{align}
where $p_B^*$ refers to the four-momentum of the tag side, $p_X^*$ is the four-momentum of the hadronic system associated with the $B \to X_u \, l \, \bar \nu_l$ candidate, and $p_l^*$ the four-momentum of the lepton candidate. For a proper $B \to X \, l \, \bar \nu_l$ decay the missing mass, $m_{\text{miss}}^2 =  p_{\text{miss}}^{*\,2}$ is required to peak at zero and it is required that $-0.5\, \text{GeV} <  m_{\text{miss}}^2 < 0.5\, \text{GeV}$. Further, a veto similar to Eq.~(\ref{eq:mmsqDstar}) is implemented. Finally, if a charged kaon or a $K_S^0$ is identified by a particle identification algorithm in the tracks of the recoiling $B$ meson, the event is rejected. The combined selection rejects $\backsim 90\%$ of the background decays that passed the preselection, while retaining about $\backsim 33\%$ of the signal decays.

The applied vetos allow the separation of the candidates into two samples: a \emph{signal enriched} sample which includes the candidates that pass all the selection criteria, and a \emph{signal depleted} sample, which contains all candidates that passed the preselection but failed at least one veto requirement. The signal depleted sample is rich in $B \to X_c \, l \, \bar \nu_l$ events and can be used to cross check the background assumptions in the signal enriched sample. Both samples are analyzed with an implicit cut in $\left| \vec p_l^{\,\,*B} \right| > 1.0 \, \text{GeV}$ unless stated otherwise and with fits in
\begin{itemize}
 \item[-] $m_X$ with a cut on $m_X < 1.55 \, \text{GeV}$ and $m_X < 1.7 \, \text{GeV}$
 \item[-] $p_X^+$ and a cut on $p_X^+ < 0.66  \, \text{GeV}$
 \item[-] two dimensions in $m_X-q^2$, and cuts on $m_X < 1.7 \, \text{GeV}$ and $q^2 > 8 \, \text{GeV}^2$ 
 \item[-] two dimensions in $m_X-q^2$, and a cut on  $\left| \vec p_l^{\,\,*B} \right| > 1.3 \, \text{GeV}$  in 
 \item[-] $\left| \vec p_l^{\,\,*B} \right|$ with a cut on $\left| \vec p_l^{\,\,*B} \right|> 1.3 \, \text{GeV}$ 
\end{itemize} 

The value of $q^2$ is reconstructed using Eq.~(\ref{eq:q2_tagged}) and the fit has two free parameters: the $B \to X_u \, l \, \bar \nu_l$ signal yield, and the summed $B \to X_c \, l \, \bar \nu_l$ and other background yield. 

The $B \to X_u \, l \, \bar \nu_l$  signal is modelled similarly as in Ref.~\cite{PhysRevLett.104.021801}, i.e. with a mix of resonant and non-resonant decays. The resonant $B \to \pi \, l \, \bar \nu_l$ decays are modelled using the model of Ref.~\cite{becpi}, with the measured model parameter of Ref.~\cite{babarpibk}. For $B \to \eta \, l \, \bar \nu_l$,  $B \to \eta' \, l \, \bar \nu_l$,  $B \to \rho \, l \, \bar \nu_l$, and $B \to \omega \, l \, \bar \nu_l$ the form factor parametrization of Refs.~\cite{Ball:hep-ph0406232,Ball:2004rg} are used. The corresponding exclusive branching fractions are fixed at the measured world-averages as calculated by Ref.~\cite{Barberio:2008fa}. The non-resonant contributions are modelled using the shape function parametrization of Ref.~\cite{1126-6708-1999-06-017}.

The $B \to X_c \, l \, \bar \nu_l$ contributions are modelled as the sum of the exclusive $B \to D \, l \, \bar \nu_l$, $B \to D^* \, l \, \bar \nu_l$, $B \to D^{**} \, l \, \bar \nu_l$, and non-resonant $B \to D^{(*)} \, \pi \, l \, \bar \nu_l$ decays using the form factor parametrizations of Refs.~\cite{Caprini:1997mu,Leibovich:1997em,Goity:1994xn}. The ratio of the resonant $B \to D^{**} \, l \, \bar \nu_l$  with respect to the summed $B \to D^{(*)} \, l \, \bar \nu_l$ plus all other background, i.e. 
\begin{align}\label{lambdaDds}
\lambda_{D^{**}} & \eq \tfrac{  \mathcal{B}(B \to D^{**} \, l \, \bar \nu_l)    } {  \mathcal{B}(B \to D^* \, l \, \nu_l) + \text{other background} }  
\end{align}
is determined from a two-dimensional fit in $q^2-m_X$ to the \emph{signal depleted} sample. The binning in the signal region was chosen to minimize the dependence on the $B \to X_u \, l \, \bar \nu_l$  signal model. The results of the one-dimensional and two-dimensional fits for the signal and background yields are depicted in Figures~\ref{babar_incl_fitresult} and \ref{babar_incl_fitresult2}. $ \left| V_{\text{ub}} \right| $ can be readily obtained from the fitted signal yields and the total inclusive $B \to X \, l \, \bar \nu_l$ branching fraction using Eq.~(\ref{eq:vub}). The authors use the predicted differential decay rates from Refs.~\cite{Lange:2005yw,Gardi:2008bb,Gambino:2007rp,adfr1,adfr2} and Table \ref{vub_babar_incl} lists the determined values of $ \left| V_{\text{ub}} \right| $ from the two-dimensional fit in $q^2$ and $m_X$ with the implicit cut of $\left| \vec p_l^{\,\,*B} \right|> 1.0 \, \text{GeV}$.

\myfigure{tbpH!}{
\vspace{-5ex}
\parbox{0.95\textwidth} { \begin{center} \includegraphics[width=0.95\textwidth]{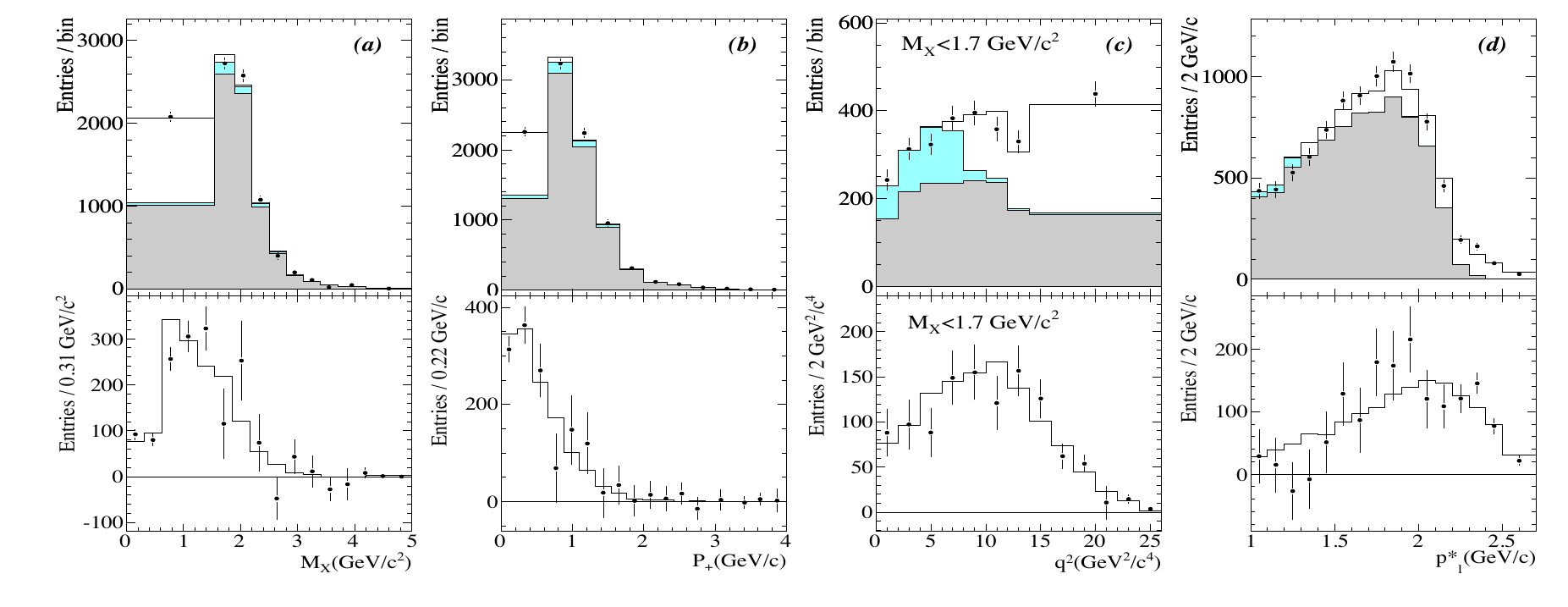}  \end{center}   }
\vspace{-2ex} 
}{ The various projections of the kinematic quantities of the one-dimensional fits to the $B \to X_u \, l \, \bar \nu_l$ and background yields are shown. Black depicts the data points, the $B \to X_u \, l \, \bar \nu_l$ signal is shown as a histogram shaded in white, $B \to X_u \, l \, \bar \nu_l$ whose actual true kinematic values are outside the selected phase-space region is shown in cyan (or light grey), and the remaining summed background corresponds to the dark grey shaded histogram: (a) $m_X$, (b) $p_X^+$, (c) $q^2$, and (d) $p_l$. The bottom row depicts the rebinned background subtracted signal yields.  \label{babar_incl_fitresult}  }

\myfigure{tbpH!}{
\vspace{-7ex}
\parbox{0.70\textwidth} { \begin{center} \includegraphics[width=0.70\textwidth]{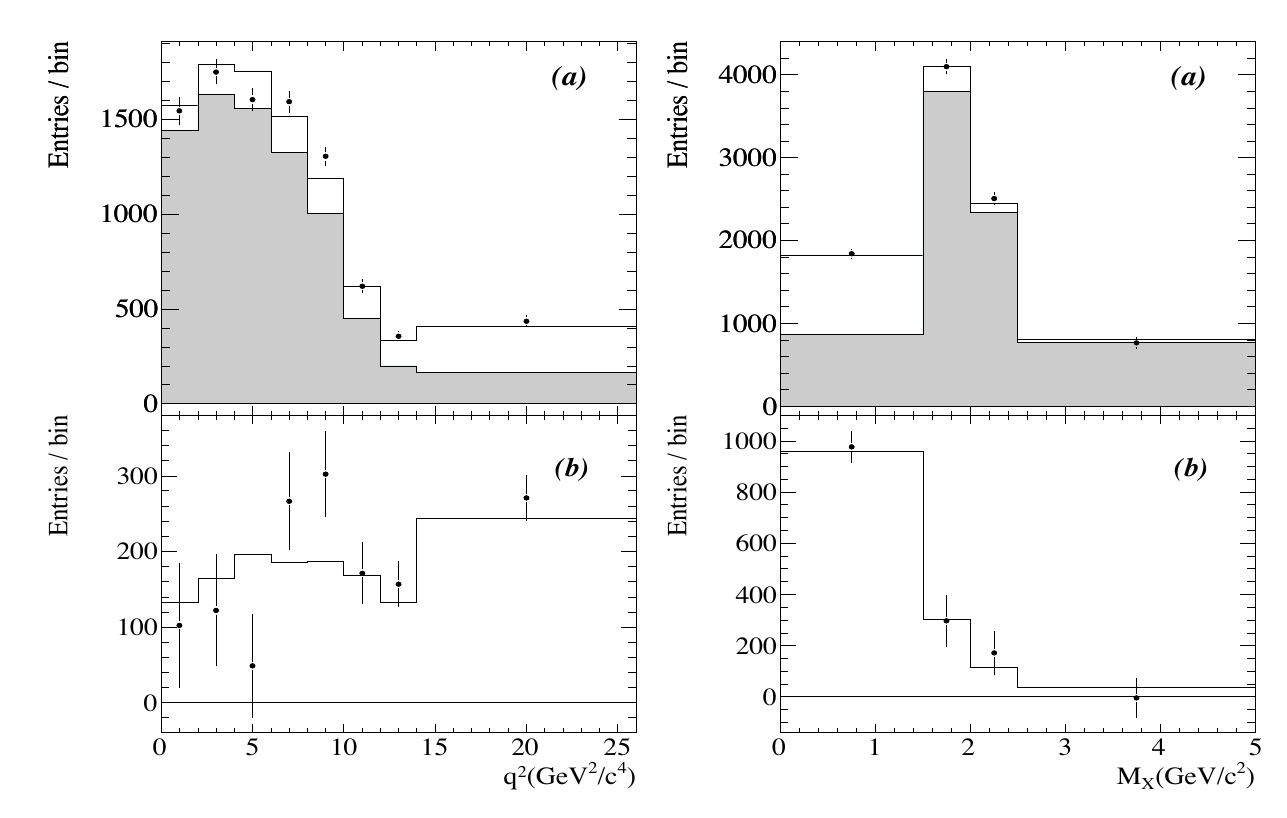}  \end{center}   }
\vspace{-2ex} 
}{ The projections of $q^2$ and $m_X$ of the two-dimensional fit in $q^2-m_X$ with an implicit cut in $\left| \vec p_l^{\,\,*B} \right| > 1.0 \, \text{GeV}$ are depicted. \label{babar_incl_fitresult2}  }

 \mytable{tbpH!}{
 \vspace{6ex}
{\footnotesize 
  \begin{tabular}{l | c c c c} 
   & BLNP \cite{Lange:2005yw} & GGOU \cite{Gambino:2007rp} & DGE \cite{Gardi:2008bb}  & ADFR \cite{adfr1,adfr2} \\  \hline
   $ \left| V_{\text{ub}} \right|  \times 10^{3}$ &   ${\bf 4.27 \pm 0.23 {}^{+0.23}_{-0.20} }$  &  ${\bf 4.29 \pm 0.24 {}^{+0.11}_{-0.14} }$ &  ${\bf 4.34 \pm 0.24 \pm 0.15 }$ & $ {\bf 4.35 \pm 0.28  {}^{+0.15}_{-0.15} }$ \\
      % $ {\bgrey {\bf 4.45 \pm 0.27 {}^{+0.24}_{-0.21} }}$  &  ${\bgrey {\bf 4.47 \pm 0.27 {}^{+0.11}_{-0.15} }}$ &  ${\bgrey {\bf 4.53 \pm 0.27 \pm 0.15 }}$ & {\bgrey -} 
  \end{tabular}\hspace{3mm}\vspace{2ex}
 }
} { The result of the determined value of  $ \left| V_{\text{ub}} \right|$ from the two-dimensional fit of Ref.~\cite{babarincl} in $q^2$ and $m_X$ with an implicit cut in $p_l > 1.0 \, \text{GeV}$ are listed. The partial decay rates were determined from the predictions of Refs.~\cite{Lange:2005yw,Gardi:2008bb,Gambino:2007rp,adfr1,adfr2}. The uncertainties are experimental and due to theory. \label{vub_babar_incl} }

%The hybrid MC is required to match the moments of the $q^2$ and $m_X$ distributions predicted by Ref.~\cite{Gambino:2007rp}.

%Using a model for the $B \to X_u \, l \, \bar \nu_l$ to determine not only the signal efficiencies, but the signal yields gives raise to some scrutiny. % 

%The $B \to X_c \, l \, \bar \nu_l$ contribution is modelled using the form factor parametrization 

%
%%%%%%%%%%%%%%%%%%%%%%%%%%%%%%%%%%%%%%%%%%%%%%%%%%%%%%%%%%%%%%%%%%%%%%%%%%%%%%%%%
%\subsection{\boldmath Study of $B \to X_u \, l \, \bar \nu_l$ from \babar}
%\label{sec:babarincl}
%%%%%%%%%%%%%%%%%%%%%%%%%%%%%%%%%%%%%%%%%%%%%%%%%%%%%%%%%%%%%%%%%%%%%%%%%%%%%%%%%
%
%The authors of Ref.~\cite{babarincl} studied $467$ million $B \bar B$ events from \babar. In order to reduce combinatorial background the $m_{\text{ES}}$ distribution of the recoil candidate is fitted. The two dimensional distribution of the hadronic mass $m_X$ of the uncharmed system and the four-momentum transfer squared $q^2$  is fitted to determine the partial branching fraction. This is done for several regions of phase-space. 

%%%%%%%%%%%%%%%%%%%%%%%%%%%%%%%%%%%%%%%%%%%%%%%%%%%%%%%%%%%%%%%%%%%%%%%%%%%%%%%%
\subsection{\boldmath Summary of inclusive results}
\label{sec:summaryincl}
%%%%%%%%%%%%%%%%%%%%%%%%%%%%%%%%%%%%%%%%%%%%%%%%%%%%%%%%%%%%%%%%%%%%%%%%%%%%%%%%

The author of Ref.~\cite{doi:10.1146/annurev.nucl.012809.104421} averaged several recent measurements of $\left| V_{\text{ub}} \right|$ using tagged or untagged approaches to determine the partial $B \to X_u \, l \, \bar \nu_l$ decay rate. Table~\ref{vub_inclusive} summarizes the results of the untagged measurements Refs.~\cite{cleo1,belle2,babar3,babar4} and compares its average with the average calculated from the two presented tagged analyses Refs.~\cite{PhysRevLett.104.021801,babarincl}. The value of $\left| V_{\text{ub}} \right|$ from both approaches are in very good agreement with each other. Further, the confidence regions of all theoretical calculations overlap. 

The good agreement of untagged and tagged results indicates that the different experimental approaches yield consistent results. Further, the good agreement between the various QCD based predictions for the inclusive decay rate indicate a sane treatment of the uncertainties related to the various approaches and approximations. However, the chosen approaches to describe the charmed background and the signal decays lead to some scrutiny: Although the used assumptions reflect our best knowledge, it also incorporates some inconsistencies which potentially could have an impact on the determined values of $\left| V_{\text{ub}} \right|$. The measured inclusive $B \to X_c \, l \, \bar \nu_l$ branching fraction is not identical to the measured exclusive contributions from $B \to D \, l \, \bar \nu_l$, $B \to D^* \, l \, \bar \nu_l$, and the measured branching fractions of the decays into the narrow and broad 1P states, $B \to D^{**} \, l \, \bar \nu_l$, i.e. a gap of about $1.4\%$ branching fraction persists. One possible explanation for the presence of this gap would be the existence of further strong decay channels of the $D^{**}$ mesons, i.e.  the $B \to D^{**} \, l \, \bar \nu_l$ branching fractions is measured only over the reconstruction of the $D^{**}$ via $D^{**} \to D^{(*)} \pi$. The presence of strong decays with an additional pion or into another light meson would allow for further contributions from $1P$ states. Such transitions, however, are not experimentally verified yet. 

Both analyses adopt a different approach to deal with this issue: Ref.~\cite{PhysRevLett.104.021801} fills the gap with broad $1P$ decays and allows the two-dimensional fit to further scale the summed background components (consisting of  $B \to X_c \, l \, \bar \nu_l$, secondary leptons and misidentified candidates). Although a systematic uncertainty is assigned (and the impact on the determined value of  $\left| V_{\text{ub}} \right|$ seems small), the authors do not specify if the adjustment of the background yields is compatible with the measured branching fractions for $B \to D \, l \, \bar \nu_l$ and $B \to D^* \, l \, \bar \nu_l$. Ref.~\cite{babarincl} fills the gap with broad and narrow $1P$ states. The fit to the signal and background yields is further allowed to scale the charmed and other background separately. In addition, the authors determine the ratio Eq.~(\ref{lambdaDds}) from their signal depleted sample, what results in an overall increase of the $1P$ contributions with respect to the $1S$ decays and the other backgrounds. The obtained value of the partial branching fraction seems fairly insensitive to the $B \to X_c \, l \, \bar \nu_l$ yields due to the charm suppressing phase-space cuts. 

Further, both analyses make the implicit assumption, that the shape of the other backgrounds (e.g. from secondary leptons and misidentified candidates) is correctly described in the MC simulation, but their corresponding yields are off. This of course is not very satisfactorily.

%However, the authors also rely on the implicit assumption that the shape of the other background components are correctly simulated.

\mytable{tbp!}{
\vspace{-3ex}
 \includegraphics[width=1.05\textwidth]{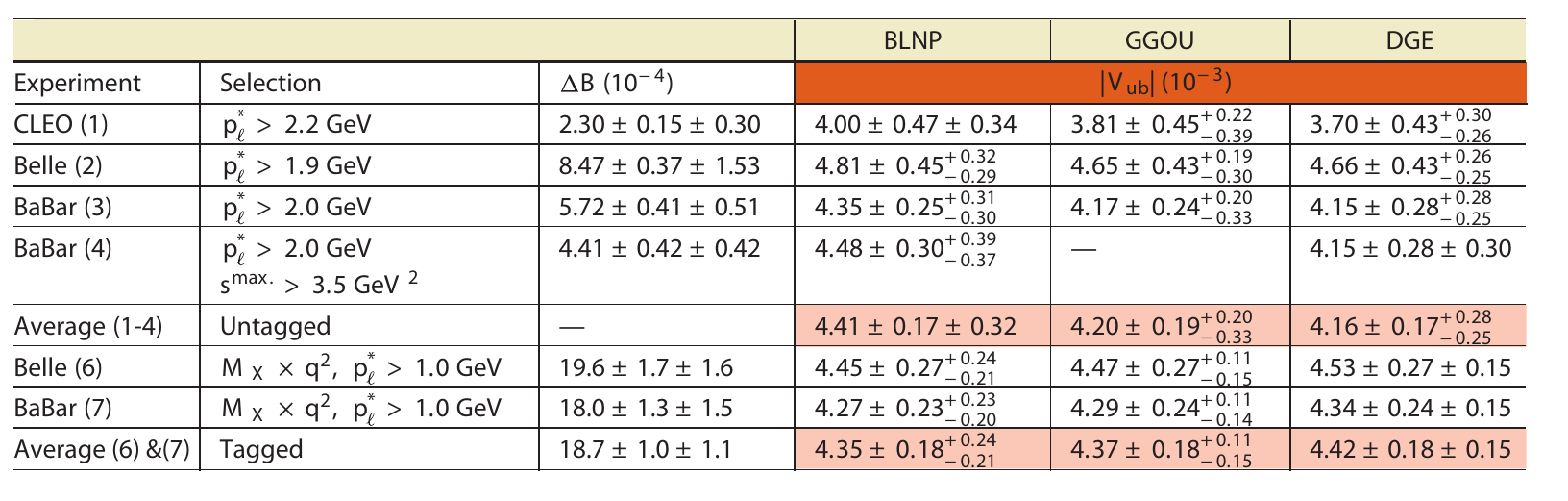}
}{ The determined value of  $\left| V_{\text{ub}} \right|$ from a selected number of untagged and the two presented tagged measurements are listed:  CLEO (1) corresponds to Ref.~\cite{cleo1}, Belle (2) to Ref.~\cite{belle2}, BaBar (3) and (4) to Refs. \cite{babar3} and \cite{babar4}. Belle (6) and BaBar (7) are the two tagged measurements presented in this talk. The stated uncertainties are experimental and from theory. Further, the Table was taken from Ref.~\cite{doi:10.1146/annurev.nucl.012809.104421}.  \label{vub_inclusive} }

\clearpage

%%%%%%%%%%%%%%%%%%%%%%%%%%%%%%%%%%%%%%%%%%%%%%%%%%%%%%%%%%%%%%%%%%%%%%%%%%%%%%%%
\section{\boldmath Exclusive determination of $\left| V_{\text{ub}} \right|$}
\label{sec:exclusive}
%%%%%%%%%%%%%%%%%%%%%%%%%%%%%%%%%%%%%%%%%%%%%%%%%%%%%%%%%%%%%%%%%%%%%%%%%%%%%%%%

In principle, many exclusive final states could be studied to determine $\left| V_{\text{ub}} \right|$, but the theoretical and experimental most promising decay channel is $B \to \pi \, l \, \bar \nu_l$: the decay signature with a single charged or uncharged pion and a high-energetic lepton allow with the combination of the known beam-energy constraints the reconstruction of the missing mass of the event, which is associated with the neutrino that eludes detection. This is often combined with a tagged approach, i.e. one of the decaying $B$ mesons is reconstructed, what allows a more clear separation of background associated with the decay of the second $B$ mesons. Such a tagging, however, results in a lower overall reconstruction efficiency, and the three analyses presented in this talk use an untagged approach.

The $B \to \pi \, l \, \bar \nu_l$ decay rate can be parametrized as a function of a few Lorentz invariant amplitudes called form factors. These can reliably be predicted from QCD sum-rules or by unquenched lattice QCD calculations and are functions of the four-momentum transfer squared of the $B$-meson to the pion system, $q^2 = \left(p_B - p_\pi\right)^2$. The value of $\left| V_{\text{ub}} \right|$ can be determined by using the measured partial branching fraction of a given $q^2$ range and the predicted partial decay rate. More recent, $\left| V_{\text{ub}} \right|$ was also determined by a simultaneous fit of the unquenched lattice calculations and the measured $q^2$ distributions. In this talk the recent measurements of Refs.~\cite{Ha:2010rf}, \cite{delAmoSanchez:2010zd}, and \cite{:2010uj} are presented. Further, the combined extracted value for  $\left| V_{\text{ub}} \right|$ of a fit to the unquenched lattice QCD results of Ref.~\cite{Bailey:2008wp} and the measured partial $B \to \pi \, l \, \bar \nu_l$ branching fractions of Refs.~\cite{Ha:2010rf,:2010uj} are presented. 

%%%%%%%%%%%%%%%%%%%%%%%%%%%%%%%%%%%%%%%%%%%%%%%%%%%%%%%%%%%%%%%%%%%%%%%%%%%%%%%%
\subsection{\boldmath Study of  $B^0 \to \pi^- \, l^+ \, \nu$  decays from \belle: Ref.~\cite{Ha:2010rf} }
\label{sec:excl3}
%%%%%%%%%%%%%%%%%%%%%%%%%%%%%%%%%%%%%%%%%%%%%%%%%%%%%%%%%%%%%%%%%%%%%%%%%%%%%%%%

The authors of Ref.~\cite{Ha:2010rf} study $B^0 \to \pi^- \, l^+ \, \nu$ from  $657 \times 10^6$ of $B \bar B$ pairs produced at the KEK-II $B$-factory and recorded by the \belle detector. The signal decay of $B^0 \to \pi^- \, l^+ \,  \nu_l$ is reconstructed from all oppositely charged leptons and pion candidates, which are required to be positively identified by a particle identification algorithm. The lepton is required to be either an electron or a muon. Both candidates are fitted to a common vertex and are required to possess a significance level of greater than $1\%$. The missing four-momentum of each event is calculated in the $\Upsilon(4S)$ rest frame by 
\begin{align}\label{eq:missingenmom}
 \left( E_{\text{miss}}^* , \, \vec p_{\text{miss}}^{\,\,*} \right) & \eq \left( \sqrt{s} - \sum_i E_i^* , \,  - \sum_i \vec p_i^{\,\,*}  \right) \, ,
\end{align}
where $\sqrt{s}$ is the average beam-energy, and the sum runs over all charged and neutral particle candidates in the event (where $(E_i^*, \vec p_i^{\,\,*})$ denotes the four-momentum of the $i^{\text{th}}$ candidate). In order to be selected an event is required to have $E_{\text{miss}}^* > 0\, \text{GeV}$. The missing three-momentum of the event is associated with the neutrino and the neutrino four-momentum is determined as $p_\nu^* \eq \left( \left| \vec p_{\text{miss}}^{\,\,*} \right| , \vec p_{\text{miss}}^{\,\,*}   \right)$ due to the higher resolution of the three-momentum reconstruction in comparison to the energy resolution. To select events compatible with the signal decay the total charge of the event is required to be small or equal to three positive or negative elementary charges. The cosine of the angle between the combined pion and lepton three-momentum and the three-momentum of the initial $B$ meson three-momentum in the $\Upsilon(4S)$ rest frame proofs to be a good discriminator to reject background decays, i.e. 
\begin{align}\label{eq:cosby}
 \cos \Theta_{BY} & \eq \tfrac{ \sqrt{s} E_{Y}^* - m_B^2 - m_Y^2  }{  2 \left| \vec p_B^{\,*} \right| \, \left| \vec p_Y^{\,*} \right| } \, ,
\end{align}
where $p_Y^* = \left( E_{Y}^*, \vec p_Y^{\,*}\right)$ denotes the summed four-momentum of the lepton and pion candidate, with mass squared $m_Y^2 = p_Y^{*\,2}$. The absolute value of the $B$-meson momentum is calculated from the beam energy as $\left| \vec p_B^{\,*} \right| = \sqrt{ s/4 - m_B^2 }$. If the lepton and pion candidate are from a signal decay, the latter cosine of the angle is required to range from $-1$ to $1$. Final state radiation is responsible for a marginal tail into negative $\cos \Theta_{BY}$ values, and resolution effects  shift a small number signal events into the region above $1$. Background decays will be shifted to large negative values, and uncorrelated lepton and pion pairs will result in a flat distribution. In order to suppress background it is required that  $ -1 < \cos \Theta_{BY} < 1$.  Further background separation can be obtained by reconstructing the beam-energy constraint mass Eq.~(\ref{eq:mes}), where the absolute value of the $B$-meson three-momentum is calculated as 
\begin{align}
 \left| \vec p_B^{\,\,*} \right|  &\eq  \left| \vec p_\pi^{\,\,*} + \vec p_l^{\,\,*} + \vec p_\nu^{\,\,*}  \right|
\end{align}
where $\vec p_\pi^{\,\,*}$ and $\vec p_l^{\,\,*}$ denote the reconstructed and measured three-momenta of the pion and lepton candidate boosted into the $\Upsilon(4S)$ rest frame. The neutrino three-momentum $\vec p_\nu^{\,\,*}$ is inferred from the missing three-momentum of the event. In addition, the energy difference Eq.~(\ref{eq:deltaE}) is calculated using
\begin{align}
E_B^* & = E_\pi^* + E_l^* + E_\nu^* \, 
\end{align}
where $E_\pi^*$, $E_l^*$, and $E_\nu^*$ denote the energy of the pion, lepton, and neutrino candidate as calculated from the reconstructed three-momenta or the missing three-momentum, respectively, boosted into the $\Upsilon(4S)$ rest frame. Candidates outside the signal regions, defined by $\left| \Delta E \right| < 1.0 \, \text{GeV}$ and $m_{bc/ES} > 5.19 \, \text{GeV} $, are rejected. To suppress further background from continuum decays, cuts on the zeroth, first, and second Fox-Wolfram moments are implemented. In addition events with an invariant mass $m_Y$ ranging from $3.07 \, \text{GeV}$ to $3.13 \, \text{GeV}$ are rejected to suppress background from misidentified $J/\psi \to \mu^+ \mu^-$ decays. The four-momentum transfer squared from the $B$ meson to the pion is reconstructed using the four-momentum of the pion candidate and four-momentum of the $B$-meson calculated from the beam-energy $\sqrt{s}$ and its known mass. The $B$ meson lies in a cone around the $Y$ system, and a weighted average over the angular orientation is taken in order to calculate $q^2$. The potential signal candidates are binned into 13 bins in $q^2$, in $16$ bins in $\Delta E$, and in $16$ bins in $m_{bc/ES}$. 

The candidates in each $q^2$ bin undergo a further preselection which is tuned to maximize the signal to background ratio in each individual bin. The used discriminating observables are the angle between the thrust axis of the $Y$ system and the thrust axis of the rest of the event; the helicity angle of the lepton-neutrino system; and the missing mass squared of the event, i.e.
\begin{align}
 m_{\text{miss}}^2 & \eq E_{\text{miss}}^{*\,2} - \left| \vec p_{\text{miss}}^{\,\,*} \right|^2 \, .
\end{align}
The absolute value of the cosine of the angle between the two thrust axes are expected to peak near one for signal events, and to be almost uniformly distributed for background decays. The cosine of the helicity angle for the lepton-neutrino system, i.e. the cosine of the angle between the lepton three-momentum and the three-momentum of the opposite $B$ meson in the rest frame of the lepton-neutrino pair, is expected to peak at $1$ due to the left-handed $V-A$ structure of the weak decay. The missing mass gives an indicator for the compatibility of the rest of the event with a decay involving a neutrino. 

The signal yields are determined by performing a two-dimensional binned maximum likelihood fit in $\Delta E$ and $m_{bc/ES}$ in the plane of the 13 $q^2$ bins, i.e. in total $13 \times 16 \times 16$ bins are considered. The probability density functions (PDF) describing the signal and the various background yields in $\Delta E$ and $m_{bc/ES}$ are obtained using MC simulations. To reduce the free parameters, the $q^2$ bins of background yields from $b \to u$ and $b \to c$ transitions are grouped in a coarser binning, e.g. four and three bins respectively. The $q^2$ distribution from continuum background is described by MC, which was reweighed to match the $q^2$ distribution of an off-resonance data sample. The continuum normalization is not a free parameter in the fit, but fixed at the scaled number of expected off-resonance events. Including the signal yields in each $q^2$ bin, there are 20 free parameters in the fit. In particular, 4 yields describe the $B \to X_u \, l \, \bar \nu_l$ background, and 3 yields the background due to $B \to X_c \, l \, \bar \nu_l$ decays. The resulting number of signal and background yields agree well with the expectations from MC simulation studies and the resulting efficiency corrected and unfolded $q^2$ spectrum is depicted in Fig.~\ref{belleexcl_q2_spectra}. 
%The leading systematic uncertainties of the measurement are due to the performance of the particle identification algorithm, the quality cut on the vertex fit probability of the fit to the pion-lepton vertex, the tracking efficiency of the detector, and the composition and modelling of the $B \to X_u \, l \, \bar \nu_l$ and $B \to X_c \, l \, \bar \nu_l$ background. 
The largest systematic uncertainties of the measurement are due to the limited knowledge of the PDF from continuum background, the $B^0 \to \pi^- \, l^+ \,  \nu_l$ signal modelling, and the background modelling from $B^0 \to \rho^- \, l^+ \,  \nu_l$ decays with $\rho^- \to \pi^- \pi^0$. Other dominant contributions to the systematic uncertainties are due to the performance of the particle identification algorithms, the sensitivity on the quality cut on the vertex fit probability, the uncertainties due to tracking efficiency, and the composition and modelling of the $B \to X_u \, l \, \bar \nu_l$ and $B \to X_c \, l \, \bar \nu_l$ backgrounds. 

The value of $\left| V_{\text{ub}} \right|$ can be calculated using Eq.~(\ref{eq:vub}) and predictions for the (partial) decay rates from theory: The unquenched lattice QCD calculations of Refs.~\cite{hpqcd,fnal} and the sum rule results Ref.~\cite{Ball:hep-ph0406232} predict the value of the form factors at several points at high $q^2$, i.e. $q^2 > 16 \, \text{GeV}^2$. Using a specific parametrization for the form factors, a prediction for the partial and total decay rate can be obtained. The values of $\left| V_{\text{ub}} \right|$ using this approach are stated in Table~\ref{tab:belle_conv}. 

This ansatz, however, potentially introduces undesired model dependence due to the specific chosen parametrization. This model dependence is somewhat minimized by only calculating  $\left| V_{\text{ub}} \right|$ within the $q^2$ range the lattice QCD or sum-rule prediction is considered reliable. This, however, is somehow unsatisfying and the authors of Ref.~\cite{:2010uj} also explore the possibility of a model-independent extraction of $\left| V_{\text{ub}} \right|$, which makes use of the pro-forma model-independent parametrization of Ref.~\cite{bpilattcombfit} and the predicted normalization from lattice QCD of Ref.~\cite{fnal}. In order to perform a combined fit of the parametrization of \cite{bpilattcombfit} and the points from \cite{fnal}, the $q^2$ spectrum, however, has to be transformed into a variable $z$, which is further discussed in Ref.~\cite{zvar}. The result of this fit is depicted in Figure~\ref{comblattdatafitzvar}. The normalization of the fit can be used to calculate  $\left| V_{\text{ub}} \right|$ and the authors of  Ref.~\cite{:2010uj} quote
\begin{align}
 \left| V_{\text{ub}} \right| & \eq \left( 3.43 \pm 0.33 \right) \times 10^{-3} \, ,
\end{align}
where the uncertainty is from both experimental and theoretical sources. 

\myfigure{tbp!}{
\vspace{-2ex}
 \includegraphics[width=0.42\textwidth]{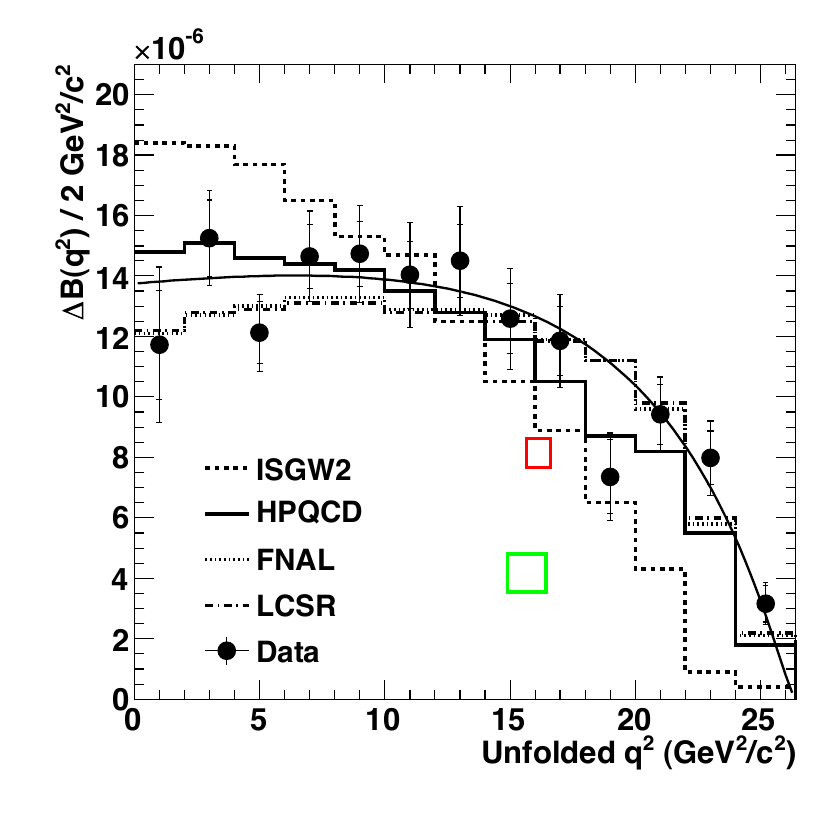}
 }{ The efficiency corrected and unfolded $q^2$ spectra of $B^0 \to \pi^- \, l^+ \, \nul$ of Ref~\cite{Ha:2010rf} is shown. The solid curve shows the result of a fit using the parametrization of Ref.~\cite{becpi}. The solid histogram shows the prediction using lattice QCD of Ref.~\cite{hpqcd}. The dotted histograms corresponds to the lattice QCD prediction of Ref.~\cite{fnal}. The dashed-dotted histogram corresponds to the QCD sum-rule prediction of \cite{Ball:hep-ph0406232}. The dashed histogram shows the quark-model prediction of Ref.~\cite{Scora:1995ty}. All histograms were scaled by $\left| V_{\text{ub}} \right|^2$ as calculated from Eq.~(\ref{eq:vub}) in the $q^2$ range the calculation is deemed reliable.  \label{belleexcl_q2_spectra} }

\mytable{tbpH!}{
\vspace{-2ex}
{\footnotesize 
  \begin{tabular}{l c c} 
                 & $q^2/ \text{GeV}^2 $ & ${\left|V_{\text{ub}} \right|} \times 10^{3}$  \vspace{1ex}   \\  \hline
HPQCD \cite{hpqcd} & $> 16$  & ${ 3.55 \pm 0.13 {}^{+ 0.62}_{-0.41}}$ \\
FNAL \cite{fnal} & $> 16$  & ${ 3.78 \pm 0.14 {}^{+ 0.65}_{-0.43}}$ \\
LCSR  \cite{Ball:hep-ph0406232}   & $< 16$  & ${ 3.64 \pm 0.11 {}^{+ 0.60}_{-0.40}}$ 
 \end{tabular}\hspace{3mm}
 }
 \vspace{3mm}
 \vspace{-1ex}
} {  The extracted values of $\left|V_{\text{ub}}\right|$ of Ref.~\cite{Ha:2010rf} obtained using the form factor predictions obtained from the lattice QCD points of Refs. \cite{hpqcd,fnal}, and the sum-rule prediction of \cite{Ball:hep-ph0406232}. The uncertainties are experimental and from theory. \label{tab:belle_conv}  }

\myfigure{tbp!}{
\vspace{-2ex}
 \includegraphics[width=0.52\textwidth]{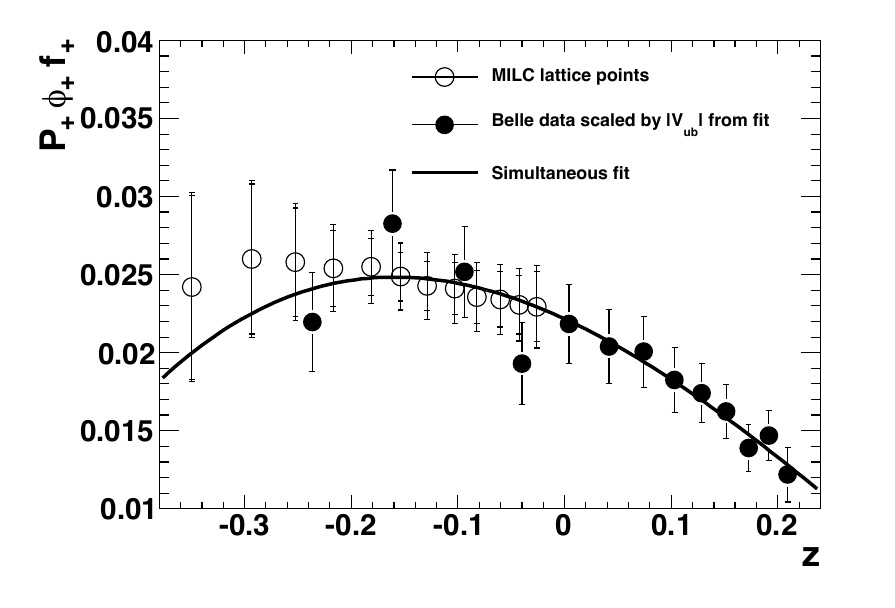}
 }{ The result of the combined lattice QCD and data points in the parametrization of Ref.~\cite{bpilattcombfit} as presented in Ref.~\cite{Ha:2010rf} is shown. The variable $z$ is defined in Ref.~\cite{zvar}. The determined normalization can be used to calculate $\left|V_{\text{ub}}\right|$. \label{comblattdatafitzvar} }

\clearpage

%%%%%%%%%%%%%%%%%%%%%%%%%%%%%%%%%%%%%%%%%%%%%%%%%%%%%%%%%%%%%%%%%%%%%%%%%%%%%%%%%
\subsection{\boldmath Study of $B \to \pi \, l \, \nul$ and $B \to \rho \, l \, \nul$ decays from \babar: Ref.~\cite{:2010uj}}
%%%%%%%%%%%%%%%%%%%%%%%%%%%%%%%%%%%%%%%%%%%%%%%%%%%%%%%%%%%%%%%%%%%%%%%%%%%%%%%%%

The authors of Ref.~\cite{:2010uj} study the decays of $B^0 \to \pi^- \, l^+ \, \nul$, $B^+ \to \pi^0 \, l^+ \, \nul$, $B^0 \, \rho^- \, l^+ \, \nul$, and $B^+ \to \rho^0 \, l^+ \, \nul$ using $377 \times 10^6$  $B \bar B$ pairs produced at the PEP-II $B$-factory and recorded by the \babar detector. Similar to Ref.~\cite{Ha:2010rf} an untagged approach is used. The identification of the four signal decays require events with at least four charged tracks, the identification of an electron or muon using an particle identification algorithm and the presence of one or more charged or neutral pions. The lepton candidates are selected from charged tracks which have at least a three-momentum of $\left| \vec p_l^{\,*} \right| > 1.0 \, \text{GeV}$ or $\left| \vec p_l^{\,*} \right| > 1.8 \, \text{GeV}$ for $B \to \pi \, l \, \bar \nu_l$ or $B \to \rho \, l \, \bar \nu_l$, respectively, in the rest frame of the $\Upsilon(4S)$. To suppress background from $e^+ \, e^- \to e^+ \, e^- (\gamma) $, the event as a whole is required to have a minimal longitudinal three-momentum such that $\zeta_z < 0.65$ with 
\begin{align}
 \zeta_z & \eq \sum_i p_i^z / \sum_i E_i^z \, ,
\end{align}
where the sum runs over all charged particles in the event, and $p_i^z$ and $E_i$ correspond to the the longitudinal three-momentum and energy measured in the laboratory frame. All tracks used for the reconstruction of the hadron candidate must be rejected by a lepton and kaon particle identification algorithm. Neutral pions are reconstructed from pairs of photons. The reconstructed $\pi^0$ is rejected if it has not at least a three-momentum of $0.2 \, \text{GeV}$. Candidates for $\rho^\pm \to \pi^\pm \pi^0$ and $\rho^0 \to \pi^+ \pi^-$ decays are required to have a two-pion mass, $m_{\pi\pi},$ within the full width of the $\rho$-meson mass, i.e. $0.65 \, \text{GeV}  < m_{\pi\pi} < 0.85  \, \text{GeV} $. To reduce combinatorial background, the three-momentum in the centre-of-mass frame of the $\Upsilon(4S)$ of one pion candidates has to exceed $0.4 \, \text{GeV} $, and further, that the second pion candidate has a three-momentum of at least $0.2 \, \text{GeV} $. Every hadron candidate is combined with one lepton candidates to form a $Y$ candidate, and a vertex fit is performed. A signal candidate is required to have at least a vertex fit significance of $0.1\%$, and a variety of other kinematic cuts on the lepton and hadron candidate kinematics are performed to reject background. The value of $\cos \Theta_{BY}$, as defined in Eq.~(\ref{eq:cosby}), is determined and in order to pass the preselection an candidate is required to have $ -1.2 < \cos \Theta_{BY} < 1.1$, what is somewhat better than the hard cut imposed in the preselection of Ref.~\cite{Ha:2010rf}, since it reduces the sensitivity due to resolution effects. The neutrino four-momentum is reconstructed from the missing energy and three-momentum of each event using Eq.~(\ref{eq:missingenmom}), but in contrast to Ref~\cite{Ha:2010rf} the missing energy and three-momentum are determined in the laboratory frame. 
%, i.e. in the rest frame of the $\Upsilon(4S)$ 
%\begin{align}
% p_\nu^* & \eq \left( E_{\text{miss}}^* \, , \vec p_{\text{miss}}^{\,\,*} \right) \, ,
%\end{align}
Further acceptance cuts are imposed to improve the reconstruction of the missing energy. For a correctly reconstructed event with a single semileptonic decay, the missing mass squared, $m_{\text{miss}}^2$, is consistent with zero, and the failure to detect one or more particles in the event results in a considerable tail towards positive values. Candidates are required to have $m_{\text{miss}}^2 / 2 E_{\text{miss}} < 2.5 \, \text{GeV}$, where $E_{\text{miss}}$ denotes the missing energy in the laboratory frame. For the signal extraction the beam-energy substituted mass, $m_{bc/ES}$, and the energy difference, $\Delta E$, are reconstructed. In contrast to Ref.~\cite{Ha:2010rf}, both variables are determined in the laboratory frame, i.e.
\begin{align}\label{eq:mes_lab}
 m_{bc/ES} & \eq \sqrt{ \tfrac{ \left(s/2 + \vec p_B \cdot \vec p_{e^+e^-}  \right)^2 }{ E_{e^+e^-}^2 } - p_B^2} \, ,
\end{align}
and
\begin{align}\label{eq:deltaE_lab}
 \Delta E & \eq \tfrac{ p_B \cdot p_{e^+e^-}  - s/2 }{ \sqrt{s}} \, ,
\end{align}
where $p_{e^+e^-} = \left( E_{e^+e^-}  , \vec p_{e^+e^-}  \right) $ denotes the measured four momentum of the colliding beam particles, and $\sqrt{s}$ is the average centre-of-mass energy of the colliding beam particles. The $B$-meson four-momentum is calculated from the measured three-momenta in the laboratory frame of the lepton and hadron candidates, and the missing three-momentum of the event associated with the neutrino, i.e. $\vec p_B = \vec p_l + \vec p_{\text{hadron}} + \vec p_\nu$ and $p_B = ( \sqrt{m_B^2 + \left|\vec p_B \right|^2} , \vec p_B )$. The $\Delta E - m_{bc/ES}$ plane is separated into a fit region, and a sideband region, both within an overall signal region, which was chosen to minimize background from $B \to X_c \, l \, \bar \nu_l$ decays.
% In order to suppress sizeable contributions from $B \to X_c \, l \, \bar \nu_l$ decays which peak close to $-0.2 \, \text{GeV}$ in $\Delta E$ the signal region is defined as
%\begin{align}
% -0.15 \, \text{GeV} &<\,\,\,\, \Delta E \,\,\,\,< 0.25 \, \text{GeV} \, , \\
% 5.225 \, \text{GeV} &<  m_{bc/ES} < 5.295 \, \text{GeV}  \, .
%\end{align}
The four-momentum transfer squared from the $B$-meson to the hadron system is calculated from the four-momentum of the lepton candidate and the missing three-momentum, as determined in the laboratory frame:
\begin{align}\label{eq:q2_babar_pirho}
  q^2 & \eq \left( p_l + \left( \left| \vec p_{\text{miss}} \right|, \,  \alpha \, \vec p_{\text{miss}}  \right) \right)^2 \, ,
\end{align}
where $\alpha = 1 - \Delta E / E_{\text{miss}} $ is a scaling parameter which improves the resolution in $q^2$.

In order to further improve the discrimination between signal and background, a neural-network technique based on a multi-layer perceptron is used, cf. Ref.~\cite{mlp}. Seven variables are used as input for three neural networks: \emph{1)} the angle between the thrust axis of the $Y$ candidate and the thrust axis of the rest of the event; \emph{2)} the summed absolute values of the three-momenta not belonging to the $Y$ candidate weighted with the cosine squared of the polar angle of the corresponding track with respect to the thrust axis calculated from all tracks not belonging to the $Y$ candidate; \emph{3)} $\cos \Theta_{BY}$; \emph{4)} $m_{\text{miss}}^2 / \left(2 E_{\text{miss}}\right)$; \emph{5)} the second normalized Fox-Wolfram moment; \emph{6)} the polar angle of the missing momentum in the laboratory frame; \emph{7)} the helicity angle of the lepton-neutrino pair. The three neural networks are trained to either reject continuum, $B \to X_c \, l \, \bar \nu_l$, and $B \to X_u \, l \, \bar \nu_l$ background using simulated signal and background decays. After the preselection and the filtering through the neural networks, the efficiency of selection background from $B \to X_c \, l \, \bar \nu_l$ is $\le 10^{-5}$. The efficiency of selecting continuum background is $< 10^{-6}$, and the efficiency of selecting other $B \to X_u \, l \, \bar \nu_l$ decays ranges from $0.3  \times 10^{-3}$ to $ 0.6 \times 10^{-3}$. The three neural networks thus suppress the background selection of the dominant $B \bar B$ and continuum background by a factor of $10^4$ and $10^5$. The efficiency of selecting a $B^0 \to \pi^- \, l^+ \, \nul$ or  $B^+ \to \pi^0 \, l^+ \, \nul$ signal decay is $1.8 \times 10^{-2}$ or $1.6 \times 10^{-2}$, respectively. The efficiency of selection a $B^0 \, \rho^- \, l^+ \, \nul$, and $B^+ \to \rho^0 \, l^+ \, \nul$ signal decay is $0.3 \times 10^{-2}$ or $0.8 \times 10^{-2}$. The rejection of the neural networks for simulated continuum events was studied with measured off-resonance data. In addition, the understanding of the neural network rejection of charmed semileptonic decays was studied with a $B \to X_c \, l \, \bar \nu_l$ enhanced sample, and a $B^0 \to D^{*-} \, l^+ \, \nu_l$ control sample. 

At this point, still several signal candidates per event are allowed and the simulation reproduces the candidate multiplicity well. In order to reduce the dependence on the simulated candidate multiplicity, only the $Y$ candidate with the highest significance from the vertex fit is retained. In case for  $B^+ \to \pi^0 \, l^+ \, \nul$ the candidate closest to the central value of the $\pi^0$ mass is selected. MC studies using these selection rules imply that in about $60\%$ of all cases the correct signal candidate is selected. 

The signal yields for $B^0 \to \pi^- \, l^+ \, \nul$, $B^+ \to \pi^0 \, l^+ \, \nul$, $B^0 \, \rho^- \, l^+ \, \nul$, and $B^+ \to \rho^0 \, l^+ \, \nul$ are determined in an extended maximum-likelihood fit to the three-dimensional $\Delta E - m_{bc/ES} - q^2$ distributions for each mode, which takes into account the limited knowledge of the signal PDFs in the MC simulation. The binning in the $\Delta E - m_{bc/ES}$ plane was chosen to optimize the background and signal shape discrimination by retaining an adequate statistics in all bins. Overall $47$ bins in $\Delta E - m_{bc/ES}$ are chosen and the $q^2$ range for $B \to \pi \, l \, \nul$ from $0$ to $26.4 \, \text{GeV}^2$ is divided into six bins, and the $q^2$ range for $B \to \rho \, l \, \nul$ from $0$ to $20.3 \, \text{GeV}^2$ is divided into three bins, resulting in a total number of $282$ and $141$ bins, respectively. In the nominal fit, all four signal yields are simultaneously determined from a fit to the measured $\Delta E - m_{bc/ES} - q^2$  distributions from all four channels, taking into account the cross feeds, e.g. the signal decay in one data sample may contribute to the background in another sample. Furthermore, the isospin relations between the branching fractions of $B^0 \to \pi^- \, l^+ \, \nul$ and $B^+ \to \pi^0 \, l^+ \, \nul$, and between  $B^0 \, \rho^- \, l^+ \, \nul$ and $B^+ \to \rho^0 \, l^+ \, \nul$, are imposed. The yields for $B \to X_u \, l \, \bar \nu_l$ background  for $B \to \pi \, l\, \bar \nu_l $ is determined in two regions, i.e. $q^2 < 20 \, \text{GeV}$, and $q^2 > 20 \, \text{GeV}^2$. For $B \to \rho \, l\, \bar \nu_l $ the $B \to X_u \, l \, \bar \nu_l$ background is fixed, due to the lack of discriminative power between the signal and $B \to X_u \, l \, \bar \nu_l$ decays. The background from other $B \bar B$ processes is separated into two yields: the dominant $B \to D^* \, l \, \bar \nu_l$ and 'other' background (consisting of the remaining $B \to X_c \, l \, \bar \nu_l$ and other $B \bar B$ processes, e.g. background producing secondary or misidentified leptons). In addition, the yields due to continuum background is a free parameter. The cross-feeds between the $B \to \pi \, l \, \nul$ and the $B \to \rho \, l \, \nul$ samples is also a free parameter of the fit. In the nominal fit, all background parameters which are not fixed, are fitted separately for each signal decay, since the different final states lead to different combinatorial backgrounds which are a-priori unrelated. The fit significance of $68.5\%$ is excellent and the resulting isospin combined $q^2$ distributions, which were unfolded and efficiency corrected, are depicted in Fig.~\ref{babarexcl1_q2_spectra}. The leading systematic uncertainties for $B \to \pi \, l \, \nul$ are due to the uncertainty in the performance of the particle identification algorithms which identify the lepton candidates, the limited knowledge on the $K_L$ spectrum and reconstruction performance, the tracking efficiency of the detector, the photon reconstruction efficiencies, the knowledge of the continuum PDF, and the modelling of the $B \to X_c \, l \, \bar \nu_l$ and $B \to X_u \, l \, \bar \nu_l$ background. For $B \to \pi \, l \, \nul$ the largest uncertainties arise from the shape function parametrization and branching fraction  used for the modelling of the $B \to X_u \, l \, \bar \nu_l$ background components, the sensitivity to the signal shape obtained from the sum-rule prediction of Ref.~\cite{Ball:2004rg}, and from the continuum background. 

\myfigure{tbp!}{
 \includegraphics[width=0.48\textwidth]{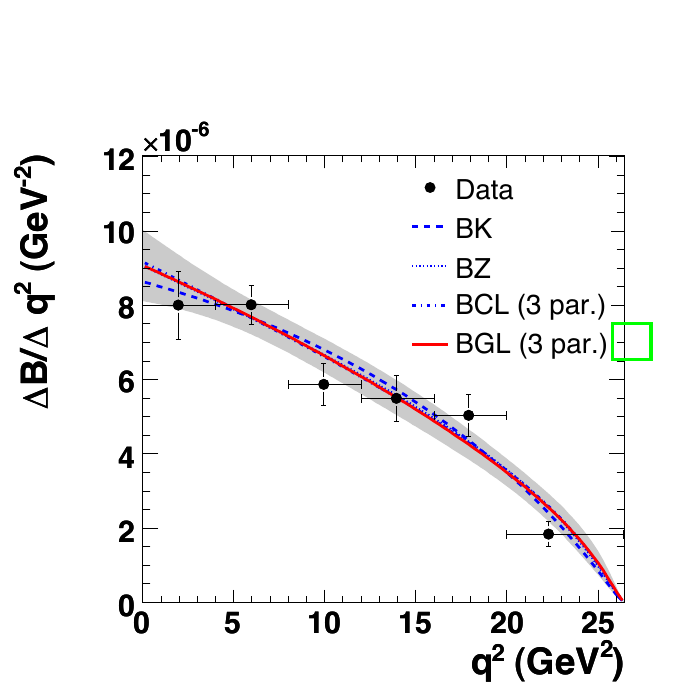}
 \includegraphics[width=0.48\textwidth]{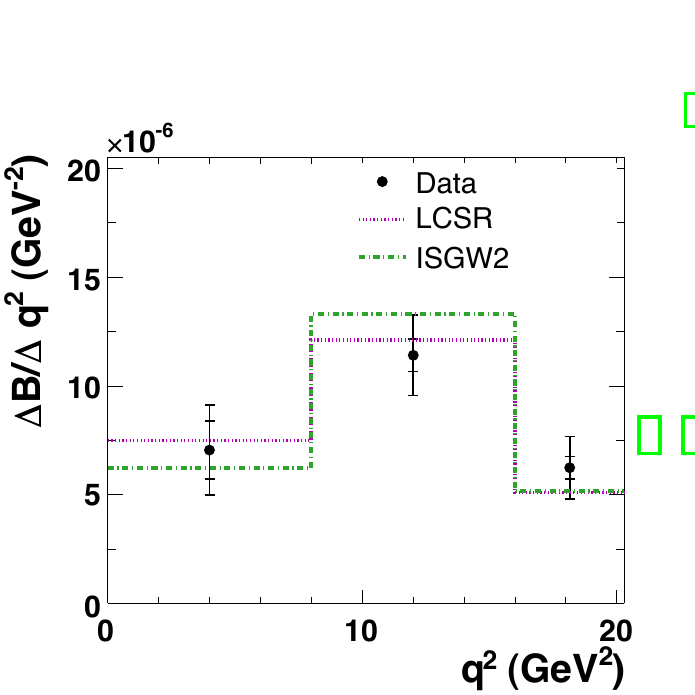}
 }{ The unfolded $q^2$ spectra of the isospin combined sample of $B\to \pi \, l \, \bar \nu_l$ (left) and $B \to \rho \, l \, \bar \nu_l$ (right) of Ref~\cite{:2010uj} are shown. Left: The blue dashed line (denoted as BK) shows the result of a fit using the parametrization of Ref.~\cite{becpi}. The blue dotted line (labeled as BZ) uses the parametrization presented in Ref~\cite{Ball:hep-ph0406232}. The blue dashed-dotted line (BCL) uses the parametrization of Ref.~\cite{bcparam}, and red line (BGL) shows the fit result of the parametrization of Refs.~\cite{bgl1,zvar}. The grey shaded region corresponds to the uncertainty of the BGL fit. Right: the predictions of Ref.~\cite{Scora:1995ty} (ISGW2), and Ref.~\cite{Ball:2004rg} (LCSR), where the ISGW2 prediction has been normalized to correspond to the total measured branching fraction, and the LCSR prediction is scaled by $\left| V_{\text{ub}} \right|^2$ calculated in the region of $q^2 < 16$. \label{babarexcl1_q2_spectra} }

The value of $\left| V_{\text{ub}} \right|$ can be calculated using Eq.~(\ref{eq:vub}) and the predictions from lattice QCD and sum rules from Refs~\cite{hpqcd,Duplancic:2008ix,Ball:hep-ph0406232,Ball:2004rg}. They are stated in Table~\ref{tab:vub_babar_excl}. 

The predictions for the differential decay rates of Refs~\cite{hpqcd,Ball:hep-ph0406232,Ball:2004rg} make use of a specific parametrization for the form factors, what introduces an undesired residual model dependence into the obtained results. The authors of Ref.~\cite{:2010uj} explore a similar approach than Ref.~\cite{Ha:2010rf} to combine the measured spectra with the unquenched lattice QCD predictions of Ref.~\cite{hpqcd,fnal} and the pro-forma model independent parametrization of Ref.~\cite{bgl1,zvar} for the form factors for a combined fit which determines $\left|V_{\text{ub}}\right|$. In contrast to the approach presented in Ref.~\cite{:2010uj}, the measured $q^2$ spectrum is not  transformed into another variable. The fit result to the full $q^2$ spectrum and four lattice QCD points of Ref~\cite{fnal} is shown in Fig.~\ref{fig:latt_fit_babar}. The fit has a good significance of $35.5\%$ and for the determined value of  $\left| V_{\text{ub}} \right|$ the authors of Ref.~\cite{:2010uj} quote
\begin{align}
 \left| V_{\text{ub}} \right| & \eq \left( 2.95 \pm 0.31 \right) \times 10^{-3} \, ,
\end{align}
where the uncertainty corresponds to the combined experimental and theory uncertainties from the lattice QCD points. 

\mytable{tbpH!}{
{\footnotesize 
  \begin{tabular}{l c c} 
    $B \to \pi \, l \, \bar \nu_l$            & $q^2/ \text{GeV}^2 $ & ${\left|V_{\text{ub}} \right|} \times 10^{3}$  \vspace{1ex}   \\  \hline
 HPQCD \cite{hpqcd} & $>16$ &   $ 3.21 \pm 0.17 {}^{+ 0.55}_{-0.36}$  \\ 
 LCSR \cite{Ball:hep-ph0406232}  &   $<16$ &   $ 3.63 \pm 0.12 {}^{+ 0.59}_{-0.40}$ \\
 LCSR \cite{Duplancic:2008ix}  &   $<12$ &   $ 3.78 \pm 0.13 {}^{+ 0.55}_{-0.40}$ \\ \hline \\
     $B \to \rho \, l \, \bar \nu_l$            & $q^2/ \text{GeV}^2 $ & ${\left|V_{\text{ub}} \right|} \times 10^{3}$  \vspace{1ex}   \\  \hline
 LCSR \cite{Ball:2004rg}  &   $<16$ &   $ 2.75 \pm 0.24$
 \end{tabular}\hspace{3mm}
 }
 \vspace{3mm}
 \vspace{-1ex}
} {  The extracted values of $\left|V_{\text{ub}}\right|$ of Ref.~\cite{:2010uj} obtained using the form factor predictions obtained from the lattice QCD points of Ref. \cite{hpqcd}, and the sum-rule predictions of \cite{Ball:hep-ph0406232,Duplancic:2008ix,Ball:2004rg}. The uncertainties are experimental and from theory. \label{tab:vub_babar_excl}  }  

\clearpage

\myfigure{tbp!}{
\vspace{-3ex}
 \includegraphics[width=0.48\textwidth]{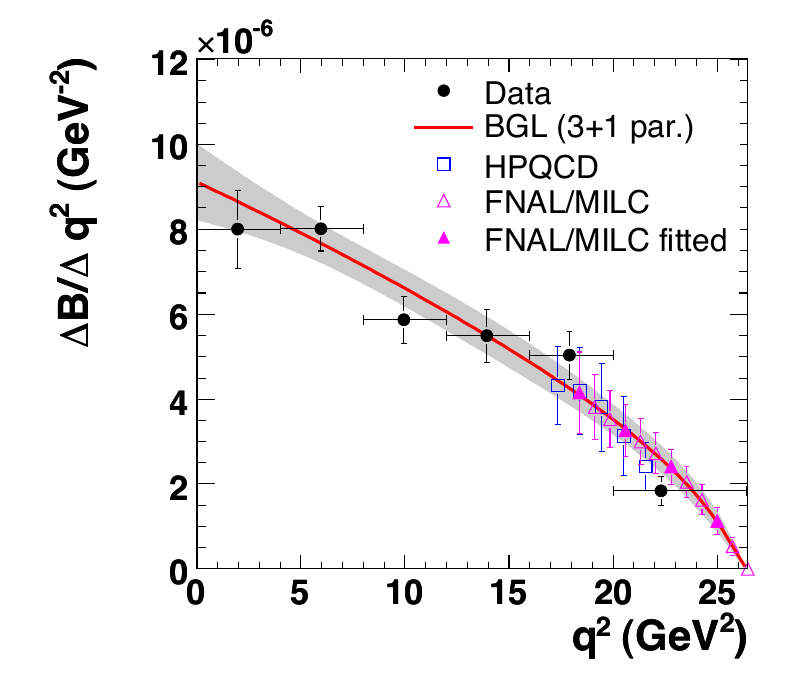}
 }{ The result of the combined lattice and data fit of Ref.~\cite{:2010uj} is shown. The four solid magenta points from Ref.~\cite{fnal} were combined with the full $q^2$ distribution to determine $\left|V_{\text{ub}}\right|$ using the form factor parametrization of Ref.~\cite{bgl1,zvar}. All lattice points were scaled by the determined value of $\left|V_{\text{ub}}\right|^2$. The grey shaded region corresponds to the uncertainty of the fit. \label{fig:latt_fit_babar} }

%%%%%%%%%%%%%%%%%%%%%%%%%%%%%%%%%%%%%%%%%%%%%%%%%%%%%%%%%%%%%%%%%%%%%%%%%%%%%%%%%
\subsection{\boldmath Study of $B^0 \to \pi^- \, l^+ \, \nu_l$ and $B^+ \to \eta^{(')} \, l^+ \, \nu_l$ decays from \babar: Ref~\cite{delAmoSanchez:2010zd}}
\label{sec:excl2}
%%%%%%%%%%%%%%%%%%%%%%%%%%%%%%%%%%%%%%%%%%%%%%%%%%%%%%%%%%%%%%%%%%%%%%%%%%%%%%%%%

The authors of Ref~\cite{delAmoSanchez:2010zd} study the decays of $B^0 \to \pi^- \, l^+ \, \nu_l$ and $B^+ \to \eta^{(')} \, l^+ \, \nu_l$  from $464 \times 10^6$  $B \bar B$ pairs produced at the PEP-II $B$-factory and recorded by the \babar detector. The analysis uses an untagged approach and a reconstructs the $B^0 \to \pi^- \, l^+ \, \nu_l$ and $B^+ \to \eta^{(')} \, l^+ \, \nu_l$  signal decay by positively identifying a track as a charged lepton by a particle identification algorithm. Electrons and muons are required to have a three-momentum greater than $0.5 \, \text{GeV}$ or  $1.0 \, \text{GeV}$, respectively. Further, for the $B^0 \to \pi^- \, l^+ \, \nu_l$ decay a positive identification from a particle identification algorithm for a pion candidate is required. The $\eta$ meson is reconstructed using the decays $\eta \to \gamma \gamma$ and $\eta \to \pi^+ \pi^- \pi^0$. The $\eta'$ is reconstructed by combining $\eta$ candidates with two charged pion candidates, i.e. through the decay chain of $\eta' \to \eta \pi^+ \pi^-$. Each possible combination of a lepton and a hadron candidate form a signal candidate, referred to as a $Y$ candidate in the following. For each $Y$ candidate a vertex fit is performed and candidates are required to have a fit significance of larger than $1\%$. Further, angular acceptance cuts on the reconstructed lepton and hadron candidates are imposed to make sure that the events were reconstructed entirely within the detector. To avoid background from $J/\psi \to \mu^+\mu^-$ decays, $Y$ candidates are required to be outside the $J/psi$ mass window. In addition, further kinematic cuts on the lepton and hadron four-momenta are imposed to improve the signal to background separation. The missing four momentum of the event is calculated using Eq.~(\ref{eq:missingenmom}). The four-momentum of the neutrino is inferred from the missing three-momentum of each event, i.e. $p_\nu = \left( \left| \vec p_{\text{miss}}^{\,\,} \right|, \, \vec p_{\text{miss}}^{\,\,}  \right)$. The four-momentum transfer squared from the $B$ meson to the hadron system is calculated as
\begin{align}
 q^2 & \eq \left( p_B - p_{\text{hadron}} \right)^2 \, ,
\end{align}
where $p_{\text{hadron}}$ denotes the four-momentum of the reconstructed hadron candidate in the laboratory frame. The $B$ meson lies in a cone around the $Y$ system, and a weighted average over the angular orientation is taken in order to calculate $q^2$. The $q^2$ spectra is further unfolded to correct for resolution effects. In contrast to Ref.~\cite{:2010uj} a cut-based approach similar to Ref.~\cite{Ha:2010rf} is used to boost the signal to background ratio for each $q^2$ bin. The following cuts are optimized: the angle between the thrust axis of the $Y$ candidate and the thrust axis of the rest of the event is optimized for each $q^2$ region;  the polar angle of the missing momentum in the laboratory frame; the missing mass squared divided by the missing energy, $m_{\text{miss}}^2 / \left(2 E_{\text{miss}}\right)$; and the helicity angle of the lepton-neutrino pair. On average, about $1.14$ candidates are observed in each selected event, and for events with multiple candidates only the candidate with the larges value of the lepton-neutrino helicity angle is kept. The signal selection efficiency varies between about $8 \%$ to $15\%$ for  $B^0 \to \pi^- \, l^+ \, \nu_l$ (depending on the $q^2$ range), and between about $1.5\%$ and $2.6\%$ for $B^+ \to \eta \, l^+ \, \nu_l$ decays in case the $\eta$ was reconstructed via two photons. In case the $\eta$ was reconstructed via three pions, the reconstruction efficiency drops to about $0.6\%$. The reconstruction efficiency of $B^+ \to \eta' \, l^+ \, \nu_l$ is also about $0.6\%$. The $B^0 \to \pi^- \, l^+ \, \nu_l$ signal selection efficiencies are considerably higher when compared to the selection efficiencies of Ref.~\cite{:2010uj}, but so are the background selection efficiencies. 

The signal yields are determined by a two-dimensional binned maximum likelihood fit in $\Delta E$ and $m_{bc/ES}$, as defined in Eqs.~(\ref{eq:deltaE_lab}) and (\ref{eq:mes_lab}). Only candidates with $\left| \Delta E \right| < 1.0 \, \text{GeV}$ and $m_{bc/ES} > 5.19 \, \text{GeV}$ are retained. The $B^0 \to \pi^- \, l^+ \, \nu_l$ background is grouped into five categories: \emph{1)} $B \to X_u \, l \, \bar \nu_l$ originating from the same $B$ as the signal candidate; \emph{2)} $B \to X_u \, l \, \bar \nu_l$ originating from the opposite $B$; \emph{3)} $B \bar B$ background originating from the same $B$; \emph{4)} $B \bar B$ background originating from the other $B$; \emph{5)} continuum background. The $B^0 \to \pi^- \, l^+ \, \nu_l$ $q^2$ distribution is divided into $12$ bins for the signal yields, and two bins for each of the five background yields. The bin widths of the background components is chosen in such a way that both bins retain adequate statistics. The PDFs for the $\Delta E$ and $m_{bc/ES}$ distributions for signal and background are determined from MC simulations. All 12 signal and two times five background yields are determined simultaneously and the resulting unfolded and efficiency corrected $q^2$ spectrum is depicted in Fig.~\ref{fig:babarexcl2_q2_spectra}. For $B^+ \to \eta \, l^+ \, \nu_l$ the $B \to X_u \, l \, \bar \nu_l$  and $B \bar B$ background categories from the opposite and same $B$ meson as the signal candidate are merged into a single yields. The $B^+ \to \eta \, l^+ \, \nu_l$ $q^2$ distribution is divided into three bins for the signal yields, and one bin for the background components. The 3 signal, the $B\bar B$ background, and the continuum yield are simultaneously determined in the fit and the background contributions from $B \to X_u \, l \, \bar \nu_l$ are fixed to the MC prediction. The $q^2$ distribution for $B^+ \to \eta \, l^+ \, \nu_l$ is as well depicted in Fig.~\ref{fig:babarexcl2_q2_spectra}. The leading systematic uncertainties in both measurements are due the modelling of the $B \to X_u \, l \, \bar \nu_l$ background and detector effects. 

\myfigure{tbp!}{
 \includegraphics[width=0.42\textwidth]{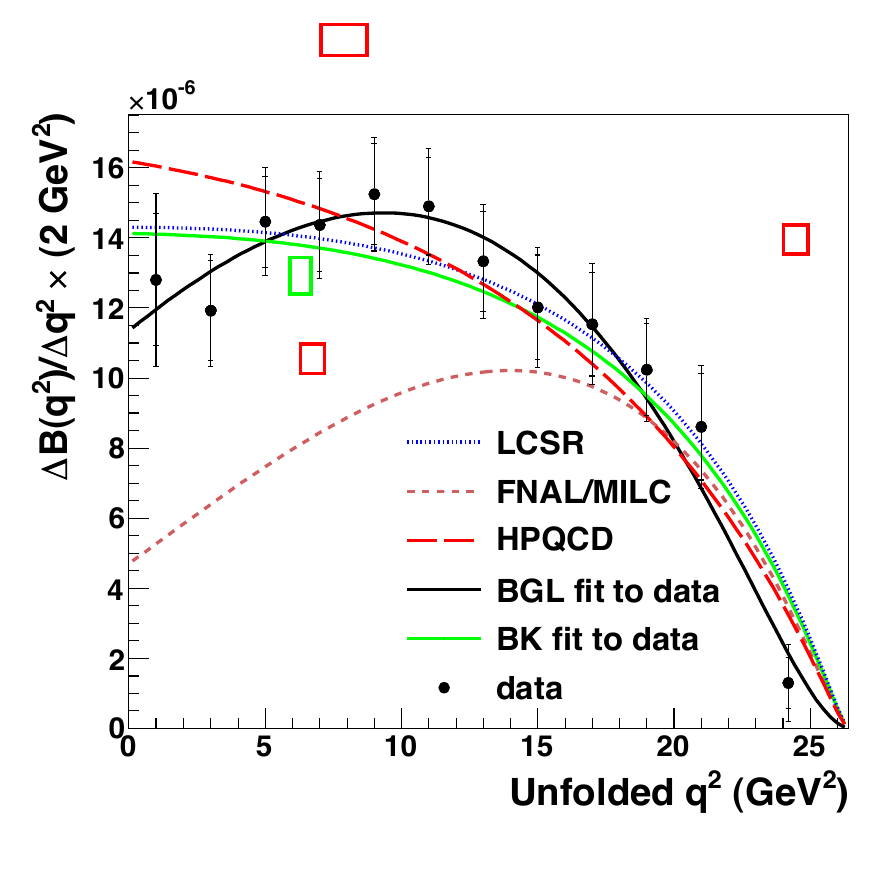}
 \includegraphics[width=0.42\textwidth]{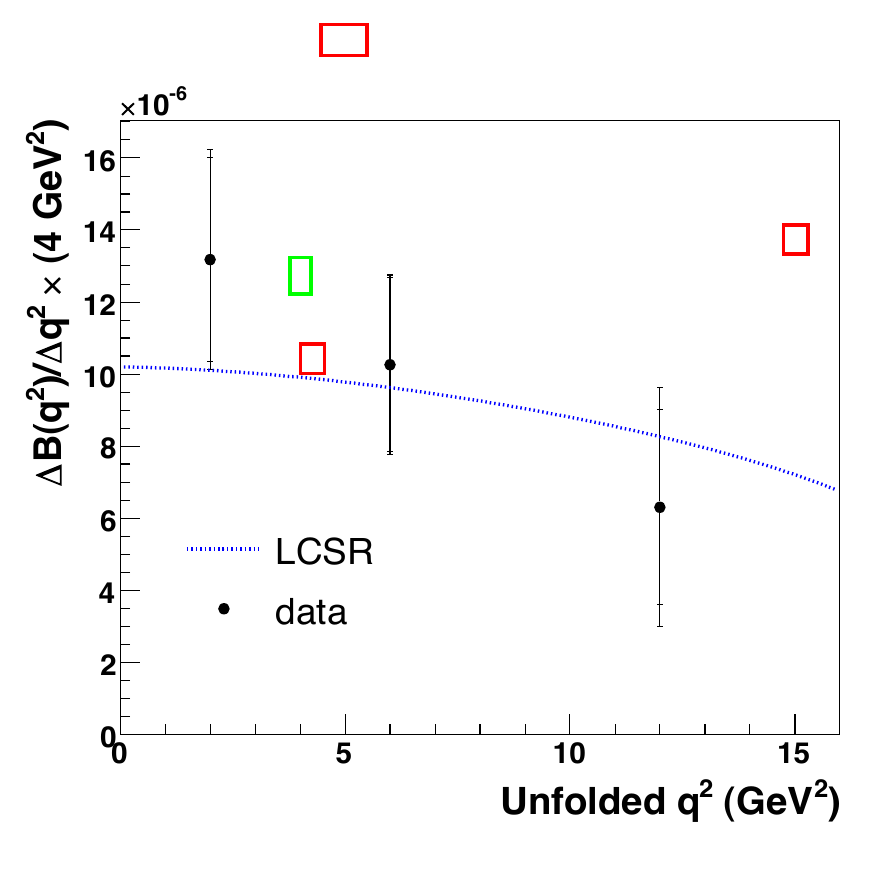}
 }{ The unfolded $q^2$ spectra of $B^0 \to \pi^- \, l^+ \, \nul$ (left) and $B^+ \to \eta \, l^+ \, \nu_l$  (right) of Ref~\cite{delAmoSanchez:2010zd} are shown. Left: the solid black and green curves show the result of a fit using the form factor parametrizations of Refs.\cite{bgl1,zvar} and \cite{becpi}. The red wide and narrow dashed lines corresponds to the prediction of Refs.~\cite{hpqcd} and \cite{fnal}, and the blue dotted line corresponds to the sum-rule prediction of Ref.~\cite{Duplancic:2008ix}. The lattice QCD and sum-rule predictions were scaled with the corresponding values of $\left|V_{\text{ub}} \right|$ squared, cf. Table~\ref{tab:vub_babar_excl2}. Right: the blue dotted line corresponds to the sum-rule prediction of Ref.~\cite{balljoneseta}, normalized to the measured branching fraction.   \label{fig:babarexcl2_q2_spectra}}
 
 The value of $\left| V_{\text{ub}} \right|$ can be calculated using Eq.~(\ref{eq:vub}) and the predictions for the partial decay rates from lattice QCD and sum rules from Refs~\cite{hpqcd,fnal,Duplancic:2008ix}. The corresponding values for  $\left| V_{\text{ub}} \right|$ are stated in Table~\ref{tab:vub_babar_excl2}.

 \mytable{tbpH!}{
{\footnotesize 
  \begin{tabular}{l c c} 
    $B \to \pi \, l \, \bar \nu_l$            & $q^2/ \text{GeV}^2 $ & ${\left|V_{\text{ub}} \right|} \times 10^{3}$  \vspace{1ex}   \\  \hline
 HPQCD \cite{hpqcd} & $>16$ &   $ 3.28 \pm 0.20 {}^{+ 0.57}_{-0.36}$  \\ 
 FNAL \cite{fnal} & $>16$ &   $ 3.14 \pm 0.18 {}^{+ 0.35}_{-0.29}$  \\ 
 LCSR \cite{Duplancic:2008ix}  &   $<12$ &   $ 3.70 \pm 0.11 {}^{+ 0.54}_{-0.39}$  
 \end{tabular}\hspace{3mm}
 }
 \vspace{3mm}
 \vspace{-1ex}
} {  The extracted values of $\left|V_{\text{ub}}\right|$ obtained using the form factor predictions obtained from the lattice QCD points of Refs.~\cite{hpqcd,fnal}, and the sum-rule prediction of \cite{Duplancic:2008ix}. The uncertainties are experimental and from theory. \label{tab:vub_babar_excl2}  }

%%%%%%%%%%%%%%%%%%%%%%%%%%%%%%%%%%%%%%%%%%%%%%%%%%%%%%%%%%%%%%%%%%%%%%%%%%%%%%%%
\subsection{\boldmath Summary of exclusive results}
\label{sec:summaryexcl}
%%%%%%%%%%%%%%%%%%%%%%%%%%%%%%%%%%%%%%%%%%%%%%%%%%%%%%%%%%%%%%%%%%%%%%%%%%%%%%%%

The author of Ref.~\cite{doi:10.1146/annurev.nucl.012809.104421} also averaged several recent measurements of the partial and total $B \to \pi \, l \, \bar \nu_l$ branching fraction using either untagged or tagged methods. Table~\ref{bf_btodpi_summary_tagged_untagged} lists the obtained averaged, which are in very good agreement in the high $q^2$ region, indicating that both experimental methods yields consistent results.  The agreement in the low $q^2$ region and the total branching fraction are less satisfactorily, hinting a systematic difference between both approaches. The background in the low $q^2$ region is dominated by $B \to X_c \, l \, \bar \nu_l$ contributions, and one possible explanation for the observed difference might be due to different model assumptions of the charmed background. 

Table~\ref{vub_btodpi_conventional} lists the obtained values of $\left| V_{\ub} \right|$ calculated from the averaged partial branching fractions using the predictions of Refs.~\cite{Ball:hep-ph0406232,hpqcd}. The large uncertainties on the sum rule and lattice QCD prediction for the differential decay rate (stripped from the CKM matrix element squared) results in a compatible result for both $q^2$ regions.  

Table~\ref{vub_btodpi_conventional_v_LQCD} compares the value for $\left| V_{\ub} \right|$ from the untagged averages of the low and high $q^2$ regions, as calculated by using the predictions of Ref.~\cite{Ball:hep-ph0406232} and Ref.~\cite{hpqcd}, with the result of the combined data and theory fits of Refs.~\cite{Ha:2010rf} and \cite{:2010uj}, which use the parametrization of Refs.~\cite{bgl1,zvar}. Needless to say, such fits fill an important gap and their implications are interesting. The determination of  $\left| V_{\ub} \right|$ from decay rates obtained by fits of model parametrization to sum rule and lattice QCD predictions neglects available information in the $q^2$ spectrum. Using a pro-forma model independent parametrization of the form factors allow a combined evaluation of the lattice QCD points at high $q^2$ region, and the entire measured $q^2$ spectrum. Furthermore, combining experimental and theoretical information to to determine the non-perturbative shape parameters of the form factors of Refs.~\cite{bgl1,zvar} minimizes the uncertainty on the predicted differential decay rate, and consequently on $\left| V_{\ub} \right|$. This is true under the premise that neither theory (e.g. through underestimating theoretical uncertainties), nor experiment (through e.g. underestimation of systematic effects or a wrong prediction for resolution effects) bias the outcome of the combined fit. These of course are non-trivial assumptions for both and lead to some scepticism towards the obtained value of $\left| V_{\ub} \right|$. However, if this premise really is violated one way or the other, the outcome of the determination of $\left| V_{\ub} \right|$ based on the fits of model functions to sum rule or lattice QCD predictions is biased as well. 

Fig.~\ref{babarbelle_excl_combfit}, finally, shows the fit of the parametrization of Refs.~\cite{bgl1,zvar} to the measured $q^2$ spectra of Refs.~\cite{Ha:2010rf} and \cite{:2010uj} and the four lattice QCD points of Ref.~\cite{fnal} (shown in bold magenta markers). The fit assumes that the systematic uncertainties of both $q^2$ spectra are $100\%$ correlated and was performed by Ref.~\cite{dingfelder}. The obtained value of $\left| V_{\ub} \right|$ is given by
\begin{align}
 \left| V_{\text{ub}} \right| & \eq \left( 3.25 \pm 0.12 \pm 0.28 \right) \times 10^{-3} \, ,
\end{align}
where the uncertainties are experimental and due to theory.

\mytable{tbp!}{
 \includegraphics[width=0.95\textwidth]{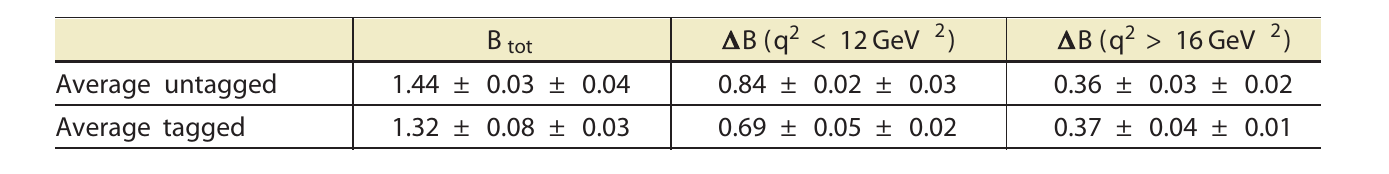}
 }{ Summary of tagged and untagged averages for the measured $B^0 \to \pi^- \, l^+ \, \nul$ branching fractions, as determined by Ref.~\cite{doi:10.1146/annurev.nucl.012809.104421}: For the tagged average the results of Refs~\cite{tagexcl1,tagexcl2,tagexcl3,tagexcl4}  were used. The untagged average was calculated from the measured total and partial branching fractions of  Refs.~\cite{Ha:2010rf,delAmoSanchez:2010zd,:2010uj}, i.e. the three presented measurements.  \label{bf_btodpi_summary_tagged_untagged} }

\mytable{tbp!}{
 \includegraphics[width=0.99\textwidth]{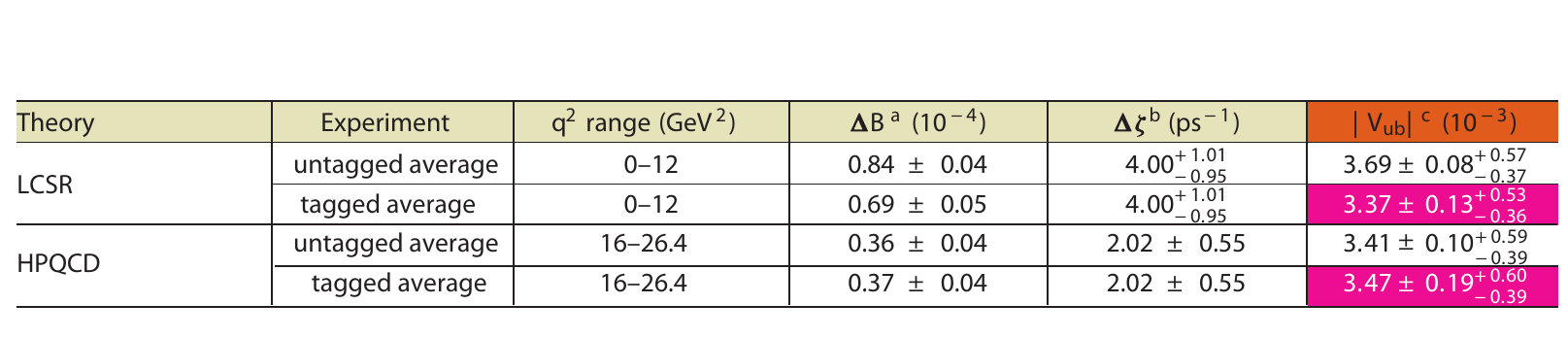}
 }{ 
  
 Summary of the tagged and untagged values of $\left| V_{\ub} \right|$ determined using the predictions of Ref.~\cite{Duplancic:2008ix} (LCSR), and Ref.~\cite{hpqcd} (HPQCD), as determined by Ref.~\cite{doi:10.1146/annurev.nucl.012809.104421}: The untagged averages correspond to the presented results Refs.~\cite{Ha:2010rf,delAmoSanchez:2010zd,:2010uj} and the tagged average are calculated from Refs~\cite{tagexcl1,tagexcl2,tagexcl3,tagexcl4}. The obtained value of $\left| V_{\ub} \right|$ from the presented measurements are highlighted in magenta.  \label{vub_btodpi_conventional} }

 \mytable{tbp!}{
 \includegraphics[width=0.88\textwidth]{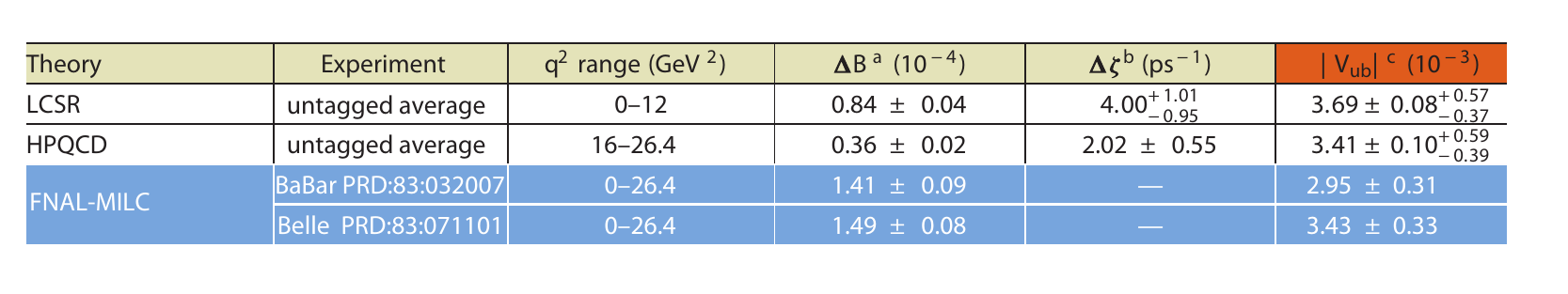}
 }{ The determined value of $\left| V_{\ub} \right|$ calculated using the predicted decay rates at low and high $q^2$, i.e. Ref.~\cite{Duplancic:2008ix} (LCSR) and Ref.~\cite{hpqcd}(HPQCD), are compared with the result of the combined fit of the whole measured $q^2$ spectrum of Refs.~\cite{Ha:2010rf} and \cite{:2010uj}, and the lattice QCD predictions of Ref.~\cite{fnal} (highlighted in blue). \label{vub_btodpi_conventional_v_LQCD} }

\myfigure{tbp!}{
 \includegraphics[width=0.82\textwidth]{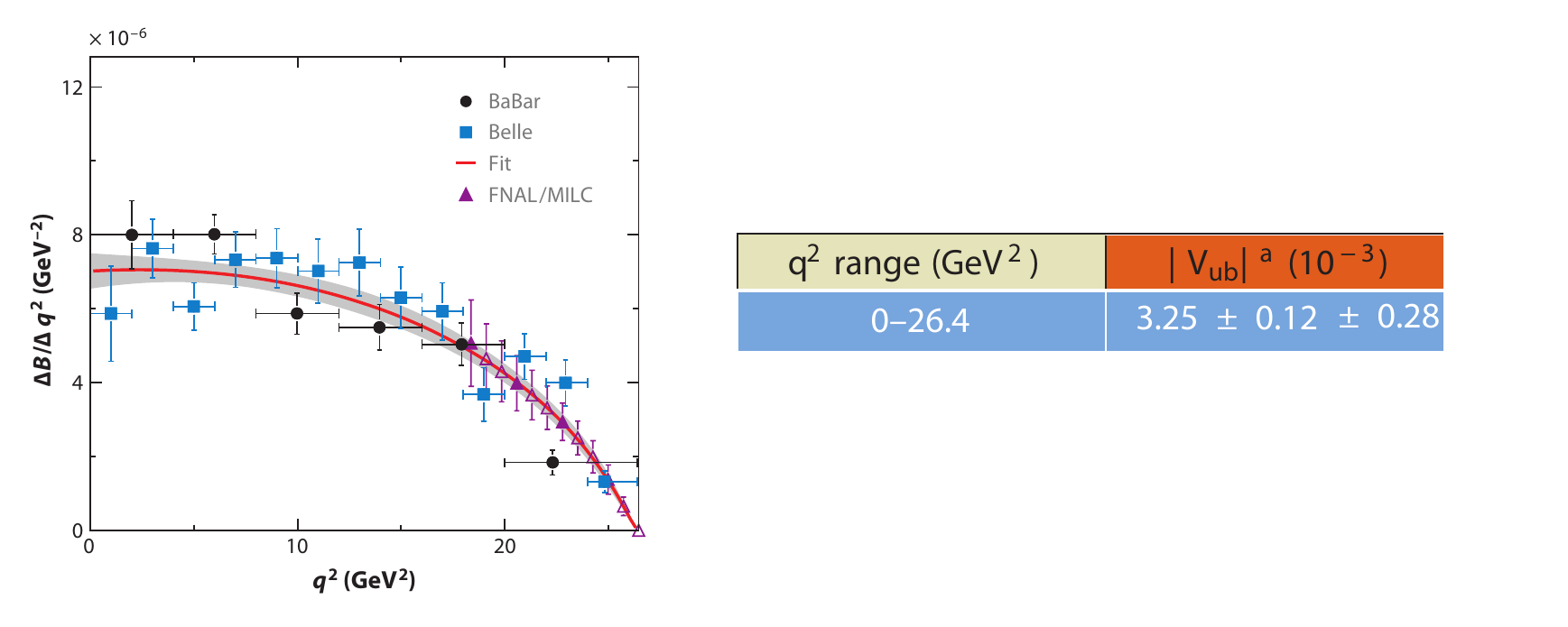}
 }{ The fit of Ref.~\cite{dingfelder} combining the experimental input of Refs.~\cite{Ha:2010rf,:2010uj} with the lattice QCD calculation of Ref.~\cite{fnal} using the parametrization of Refs.~\cite{bgl1,zvar} is shown. The lattice QCD points used in the fit are shown as filled magenta triangles and are scaled by the obtained value of $\left| V_{\text{ub}} \right|$ squared.   \label{babarbelle_excl_combfit} }

%%%%%%%%%%%%%%%%%%%%%%%%%%%%%%%%%%%%%%%%%%%%%%%%%%%%%%%%%%%%%%%%%%%%%%%%%%%%%%%%
\section{\boldmath Conclusions}
\label{sec:conclusions}
%%%%%%%%%%%%%%%%%%%%%%%%%%%%%%%%%%%%%%%%%%%%%%%%%%%%%%%%%%%%%%%%%%%%%%%%%%%%%%%%

Comparing the averaged tagged inclusive result of $\left| V_{\text{ub}} \right|$ in Table~\ref{vub_inclusive}, e.g. using the BLNP prediction, 
\begin{align}\label{eq:vub1}
 \left| V_{\text{ub}} \right| & \eq \left( 4.35 \pm 0.19 {}^{+0.24}_{-0.21} \right) \times 10^{-3} \, ,
\end{align}
with the combined data and lattice QCD fit results using Refs.~\cite{Ha:2010rf,:2010uj} and Ref.~\cite{fnal}, i.e. 
\begin{align}\label{eq:vub2}
 \left| V_{\text{ub}} \right| & \eq \left( 3.25 \pm 0.12 \pm 0.28 \right) \times 10^{-3} \, ,
\end{align}
illustrate the tension between the values of $\left| V_{\text{ub}} \right|$ obtained from inclusive and exclusive approaches: Both values differ by about $\backsim 2.5\, \sigma$, if considered being completely uncorrelated. The source of this tension is a long standing issue and interesting due to the fact, that the theoretical and experimental methods of both approaches for the determination of the partial branching fraction and $\left| V_{\text{ub}} \right|$ are often considered independent. Recently, the author of Ref.~\cite{Crivellin:2009sd} (see also Ref.~\cite{Buras:2010pz}) proposed that the presence of right-handed currents could bring the results obtained from inclusive and exclusive results into perfect agreement. Fig.~\ref{rhscan} depict the determined values $ \left| V_{\text{ub}} \right| $ quoted in Eqs.~(\ref{eq:vub1}) and (\ref{eq:vub2}) as a function of the right-handed admixture $\epsilon_R$, as proposed in Ref.~\cite{Crivellin:2009sd}: Both results overlap with an admixture of about $\epsilon_R \backsim -0.20$. A direct search for right-handed contributions can be realized in decays with a non-trivial $V-A$ and $V+A$ admixture (e.g. $B \to \rho \, l \, \bar \nu_l$). 

Another potential explanations for the tension between the inclusive and exclusive values of $\left| V_{\text{ub}} \right|$ might lie in the model assumptions for the shape function. Their uncertainty estimates raise some flags (see e.g. Ref.~\cite{frankstalk}) and a model independent determination of the shape function would be very desirable. The authors of Ref.~\cite{Bernlochner:2011di} reported some recent progress by determining the shape function with absorbed $1/m_b$ corrections from $B \to X_s \, \gamma$ decays using the proposed methods of Ref~\cite{Ligeti:2008ac}. Fig.~\ref{simba} shows the determined functional form.

\myfigure{tbp!}{
 \includegraphics[width=0.62\textwidth]{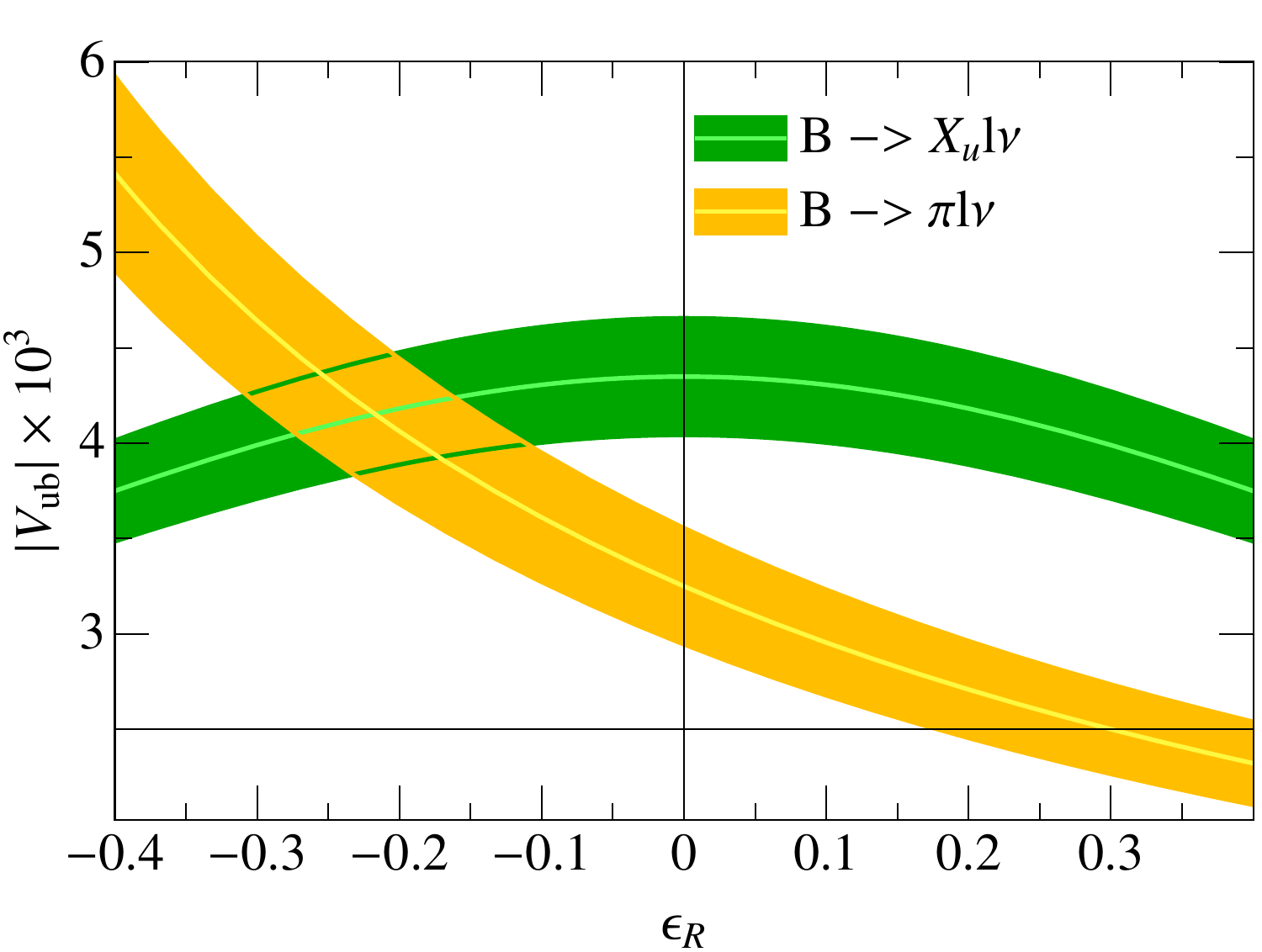}
 }{ The averaged value of $\left| V_{\text{ub}} \right|$ from the BLNP result (green) and the combined data and theory fit to Refs.~\cite{Ha:2010rf,:2010uj} and Ref.~\cite{fnal} (yellow) are shown as a function of the right-handed admixture $\epsilon_R$.  \label{rhscan} }

\myfigure{tbp!}{
 \includegraphics[width=0.62\textwidth]{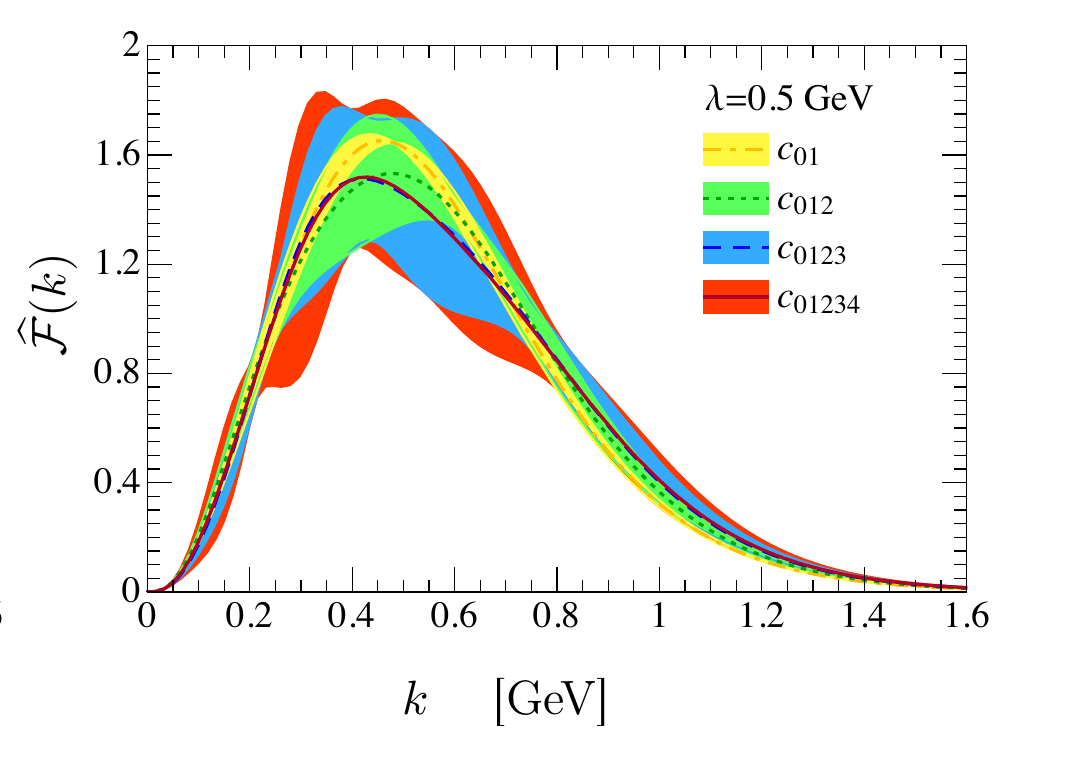}
 }{  The $B \to X_s \gamma$ shape function from Ref.~\cite{Bernlochner:2011di} is shown. It was obtained by a fit to three measured $B \to X_s \, \gamma$ photon energy spectra and a pro-forma model-independent expansion of the shape function into a set of orthogonal basis functions. The non-perturbative physics of the shape function is described by the expansion coefficients and the different coloured curves and confidence regions correspond to the fit with two ($c_{01}$), three ($c_{012}$), four ($c_{0123}$), and five ($c_{01234}$) basis functions. The confidence region cover the experimental uncertainties only.   \label{simba} }

The situation of the various different approaches of modelling the $B \to X_c \, l \, \bar \nu_l$ background is also somewhat unsatisfying. Even after a decade of having two $B$ factories running, we still don't know what exactly consist more than $\backsim 10\%$ of the total $B \to X_c \, l \, \bar \nu_l$ branching fraction. Although many analyses reject or suppress these background components considerably, this lack of knowledge limits our understanding of inclusive and exclusive $B \to X_u \, l \, \bar \nu_l$ decays. The precise determination of the charmed decays should be a priority to everyone who wants to measure charmless decays below the endpoint of the lepton spectrum, in the low $q^2$ region, or at high $m_X$. At the time being, it is this poor understanding of higher charmed resonances that pushes inclusive analyses into regions of phase-space where shape function effects play an important role and maybe the discrepancy between the inclusive and exclusive values of $ \left| V_{\text{ub}} \right| $ does not hint towards new physics, but just poorly understood QCD effects.

With potentially two super $B$-factories being built, we hopefully will be able to investigate and eventually resolve this conundrum. 

\clearpage

\bibliographystyle{plain}
\bibliography{fpcp}

\end{document}